\documentclass[aps,pre,twocolumn,groupedaddress,floatfix,superscriptaddress]{revtex4}
\usepackage{dcolumn}
\usepackage{amsmath}
\usepackage{amssymb}
\usepackage{graphicx}
\usepackage{bm}
\usepackage[T1]{fontenc}
\usepackage{color}
\usepackage{url}
\usepackage[bf]{subfigure}
\usepackage{rotating}
\usepackage{soul}
\usepackage{scalefnt}
\usepackage{hyperref}

\usepackage{perpage} %the perpage package
\MakePerPage{footnote}

\usepackage{color}
\definecolor{redcolor}{rgb}{1.0,0.,0.}
\definecolor{bluecolor}{rgb}{0.,0.,1.0}

\begin{document}

%\title{Knowledge modeling as true self avoiding walk in complex networks}
\title{Knowledge Acquisition: A Complex Networks Approach}
%\shorttitle{Title} %Insert here a short version of the title if it exceeds 70
%characters

%Definir certo a ordem dos autores
\author{Henrique Ferraz de Arruda}
\email{h.f.arruda@gmail.com}
\affiliation{Institute of Mathematics and Computer Science at S\~ao Carlos,
University of S\~ao Paulo, PO Box 668,  13560-970, S\~ao Carlos, SP, Brazil. }

\author{Filipi Nascimento Silva}
\email{filipinascimento@gmail.com}
\affiliation{S\~ao Carlos Institute of Physics, University of S\~ao Paulo, PO Box 369,  13560-970, S\~ao Carlos, SP, Brazil}

\author{Luciano da Fontoura Costa}
\email{ldfcosta@gmail.com}
\affiliation{S\~ao Carlos Institute of Physics, University of S\~ao Paulo, PO Box 369,  13560-970, S\~ao Carlos, SP, Brazil}

\author{Diego Raphael Amancio}
\email{diegoraphael@gmail.com}
\affiliation{Institute of Mathematics and Computer Science at S\~ao Carlos,
University of S\~ao Paulo, PO Box 668,  13560-970, S\~ao Carlos, SP, Brazil. }

\begin{abstract}

Complex networks have been found to provide a good representation of the structure of knowledge, as understood in terms of discoverable concepts and their relationships. In this context, the discovery process can be modeled as agents walking in a knowledge space. Recent studies proposed more realistic dynamics, including the possibility of agents being influenced by others with higher visibility or by their own memory. However, rather than dealing with these two concepts separately, as previously approached, in this study we propose a multi-agent random walk model for knowledge acquisition that incorporates both concepts. More specifically, we employed the true self avoiding walk alongside a new dynamics based on jumps, in which agents are attracted by the influence of others. That was achieved by using a L\'evy flight influenced by a field of attraction emanating from the agents. In order to evaluate our approach, we use a set of network models and two real networks, one generated from Wikipedia and another from the Web of Science. The results were analyzed globally and by regions.  In the global analysis, we found that most of the dynamics parameters do not significantly affect the discovery dynamics. The local analysis revealed a substantial difference of performance depending on the network regions where the dynamics are occurring. In particular, the dynamics at the core of networks tend to be more effective. The choice of the dynamics parameters also had no significant impact to the acquisition performance for the considered knowledge networks, even at the local scale.
\end{abstract}

\maketitle

\setcounter{secnumdepth}{1}

\section{Introduction}

Understanding how science works and evolves has become an important subject of study over the last few years. Science itself can be regarded as being a complex system, which can be approached through concepts from many disciplines such as physics, statistics, linguistics, and information science.  The knowledge achieved by humanity can be understood as a subset of the knowledge in nature. Moreover, the knowledge is constantly evolving and growing with new discoveries. Such evolution have been studied, for instance in \cite{foster2015tradition,cokol2005emergent}. In particular, scientometry~\cite{leydesdorff2009global} emerged as a new research area investigating how science evolves. Among such studies are those related to the modeling of the discovery process~\cite{doi:10.1093/comnet/cnu003,Pareschi20130396,batista2010knowledge}, which are, in general, based on investigations of the structure and dynamics existing in a knowledge space. Such dynamics usually involves researchers acquiring information while exchanging information. Furthermore, the discovery process must take into account how knowledge is organized~\cite{PhysRevE.88.012814,leydesdorff2009global,silva2011investigating,silva2016using}.

The study of how knowledge acquisition takes place allow us to identify the effects influencing the efficiency of such a dynamics and, consequently, to better understand how they can improve or constrain the process of learning. An example of such study was conducted by Silva \emph{et al}~\cite{silva2010identifying}, in which it was shown that discoveries in mathematics, in particular theorems, are more likely to be incorporated at the borders of the respective knowledge space.  In order to understand the knowledge acquisition dynamics, some key elements should be considered: (i)~how researchers choose their own research topic; (ii)~the way in which they spread the obtained results; (iii)~how fast the information spreads; (iv)~the visibility of specific researchers and (v)~the organization of the knowledge itself. Some of these elements were previously considered in literature, such as a model of spreading ideas in a group of multiple agents~\cite{burridge2016infrequent}. Other studies addressed techniques to optimize the exploration of a knowledge space~\cite{kim2016network,lopez2012model,rzhetsky2015choosing}. Another important aspect of interest is knowing how the interactions among researchers impacts the discovery process. For instance, what would be the effects implied in case researchers were strongly influenced by their more famous colleagues.

Many works tackled the characterization and modeling of science by using complex networks (e.g.~\cite{da2006learning,silva2010identifying,batista2010knowledge,gonzalez2011threshold,boyer2014random,forsman2014extending,rzhetsky2015choosing,guan2017impact}). Typically, each node represents a concept while the edges stand for the relationship between them. The dynamics of learning in such networks can be modeled as random walks, in which an agent randomly moves along the network through its edges~\cite{da2006learning,boyer2014random}. In this context, the nodes already visited by an agent represent the already learnt concepts. For example, in the methodology proposed by Costa~\cite{da2006learning}, an agent performs a random walk along a multilayer network, which represents the hierarchy of the knowledge structure. Additionally, the movement among layers is only permitted if a certain milestone of learned concepts is reached. Other characteristics can be incorporated in the random walk, such as memory, in which the agent tends to move to places already known in a network~\cite{boyer2014random}. Apart from single agent dynamics, some papers studied the \emph{collective discovery} process, in which the interactions among multiple researchers are taken into consideration~\cite{batista2010knowledge,gonzalez2011threshold,forsman2014extending,rzhetsky2015choosing}. For example, Forsman \emph{et al}~\cite{forsman2014extending} suggests that social and academic networks, as well as their dynamics, are influenced by the relationships among university-level students.

In perspective to the previous studies, we propose a model for knowledge acquisition that incorporates two main aspects typically approached separately: knowledge memory and multiple agents. The memory aspect is encompassed into the proposed dynamics by an specific type of random walk, the \emph{true self-avoiding walk} (TSAW), which was found to be one of the most efficient models to explore networks~\cite{kim2016network}. In the TSAW dynamics, the agent tends to avoid passing through already visited nodes. The dynamics is based on a set of a random walkers that simultaneously explore the network. Aside from these two aspects, we also incorporate into the dynamics the possibility of a researcher to change its research focus based on discoveries made by others. In particular, the agents have a probability of performing a jump, which can be understood as a long-range travel across the network~\cite{estrada2016random}. In the analyzes made by Foster~\emph{et~al}~\cite{foster2015tradition}, the jump dynamics was found to be related to the risk of research. On the other hand, such kind of risk can produce innovation with high impact in science.

%******************************************

In this paper, we perform knowledge dynamics for a set of network models and two real networks: a citation~\cite{silva2016using} and a Wikipedia~\cite{silva2011investigating} network. In both networks, the nodes represent articles and the edges, the citations among them. In particular, for the Wikipedia network, an edge exists between two nodes if there is a hyperlink connecting their corresponding articles.

Bearing in mind that our approach simulates collective discovery, we measure the performance in terms of the fraction of the total number of explored nodes by the agents after a certain number of iterations. Through our approach, we seek to address some important questions regarding the collective discovery process, such as: if frequent changes of the area of study by researchers can positively contribute to the performance in the discovery process; investigating influence of researchers with very high visibility to the dynamics; how important is the overall organization of knowledge to the dynamics; and how the dynamics behaves in specific regions of a network (e.g. its borders).

The current paper is organized as follows: Section~II describes the basic concepts related to the particularities of the adopted random walk model. In Section~III, we present, in detail, our model for knowledge acquisition as well as the description of the used dataset. Section~IV describes the results obtained from the proposed methodology. Finally, Section~V provides the conclusions and suggestions for further work.

\section{True self-avoiding random walk}
The problem of random walks was initially proposed by Karl Pearson, in 1905~\cite{pearson1905problem}, in order to study the dynamics of mosquito swarms in forests. Pearson assumed a mosquito as an agent under a dynamics based on the walking of a drunkard. Such a process starts with the agent placed at a point $O$ along a plane. Next, it moves straight ahead $l$ yards, then, it turns through a random angle and moves another $l$ yards in a straight line. This process is repeated $n$ times. In a short letter to \emph{Nature}, Pearson asked for help on solving the problem of devising the probability distribution $P(d, n)$ of finding such an agent at a certain distance $d$ from $O$ after a high amount of $n$ iterations. The solution was found by Lord Rayleigh~\cite{rayleigh1905problem}, who showed that $P(d, n) \approx {2 d \over n} e^{-d^2/n}$ as $n\to \infty$, a relationship that is statistically similar to the behavior of the diffusion dynamics. The established link between these two concepts (random walks and diffusion), paved the way to the development of the theoretical approaches to understand the characteristics of matter, such as the concept of Brownian motion~\cite{horvath2012diffusive}, leading to the discovering of several macroscopic characteristics and the modeling of many real-world phenomena.

More recently, some variants of the random walk dynamics were introduced, which were employed for modeling of many systems in a wide range of disciplines, including the analysis of insect movements~\cite{kareiva1983analyzing}, modeling of neural activity~\cite{gerstein1964random}, and other biological processes~\cite{codling2008random}. Subsequently, random walks were applied in conjunction with complex networks~\cite{costa2011analyzing,da2007exploring,noh2004random}, which have been used to represent several real-world complex systems. Examples of such developments are: finding influential spreaders in rumor propagation~\cite{de2014role}, characterizing proteins~\cite{rodrigues2009comparison}, text mining~\cite{mihalcea2006random} and knowledge representation~\cite{leydesdorff2009global, silva2013quantifying}.

Among the proposed variations of random walk dynamics on networks, there are those based on self-avoiding random walks (SAW)~\cite{herrero2005self,herrero2003self}. In such a walking dynamics, each agent behaves identically as it would do in a traditional random walk; however, the same agent is not allowed to return to the same node.
Because at each iteration the agent always visit a new node, this dynamic is particularly suitable for optimally exploring a network with no \emph{a priori} information of its global structure. By definition, paths generated from SAWs are finite in length~\cite{tishby2016distribution}, thus a path restart mechanism is needed for exploring the whole network.
A drawback associated with SAWs is the creation of discontinuous paths, as a consequence of the path restart mechanism.
To overcome this pitfall, variations of the self-avoiding random walk have been proposed, including the \emph{true self-avoiding walk} (TSAW)~\cite{kim2016network,lam1984true,stella1984series,amit1983asymptotic}. In such dynamics, also known as \emph{myopic walk}, the probability of visiting a neighbor is higher if it has not already been frequently accessed.
Note that the TSAW resembles the SAW dynamics, with the difference of not being fully restrictive with regard to neighbors already accessed in the previous visits.

In a typical TSAW dynamics, the next node $v^{(t+1)}$ to be taken by an agent must lie in its current neighborhood $v_i \in N(v^{(t)})$. The selection of $v^{(t+1)}$ is conditioned to a probability $P_i$ that diminishes exponentially with the number of times $v_i$ have already been visited, $f(v_i)$. Thus, the probability $P_i$ is proportional to
\begin{equation}	
	w_i = \alpha^{-f(v_i)},
\end{equation}
where $\alpha$  is a constant. Given $w_i$, the probability $P_i$ of visiting
the next node $v_i$ can be computed as
\begin{equation}
\label{eq:tsawprobability}
P_i = { w_i \over \sum\limits_{v_j \in N(v^{(t)})} w_j },
\end{equation}
Henceforth, $\alpha=2$.

Since both TSAW and SAW dynamics must maintain a memory of the path already taken by the agent, the analytical study of such both processes becomes very complex. The TSAW dynamics; nonetheless, is more analytically tractable than SAW because the former does not depend on a path restart mechanism~\cite{lopez2012model}. Recently, Kim~\emph{et al}~\cite{kim2016network} showed that the TSAW dynamics is an efficient way to explore complex networks with agents having only the knowledge of the local structure of the network.

\section{Knowledge acquisition modeling}

The process of collective discovery can be modeled as a population of agents exploring the knowledge space under certain dynamics, which should also incorporate the interactions among the agents. In this context, the knowledge space can be represented by a complex network~\cite{leydesdorff2009global, silva2013quantifying} where potential discoveries are indicated by nodes, while relationships between such discoveries are represented as edges. Examples of such networks include citation or co-citation networks~\cite{Ren20123533,PhysRevE.88.012814,Amancio2012427,silva2016using,scientometria}, Wikipedia networks~\cite{Ibrahim201721,silva2011investigating}, and learning objects~\cite{korytkowski2007creating}.

\begin{figure*}[!htbp]
 \centering
 \includegraphics[width=0.9\linewidth]{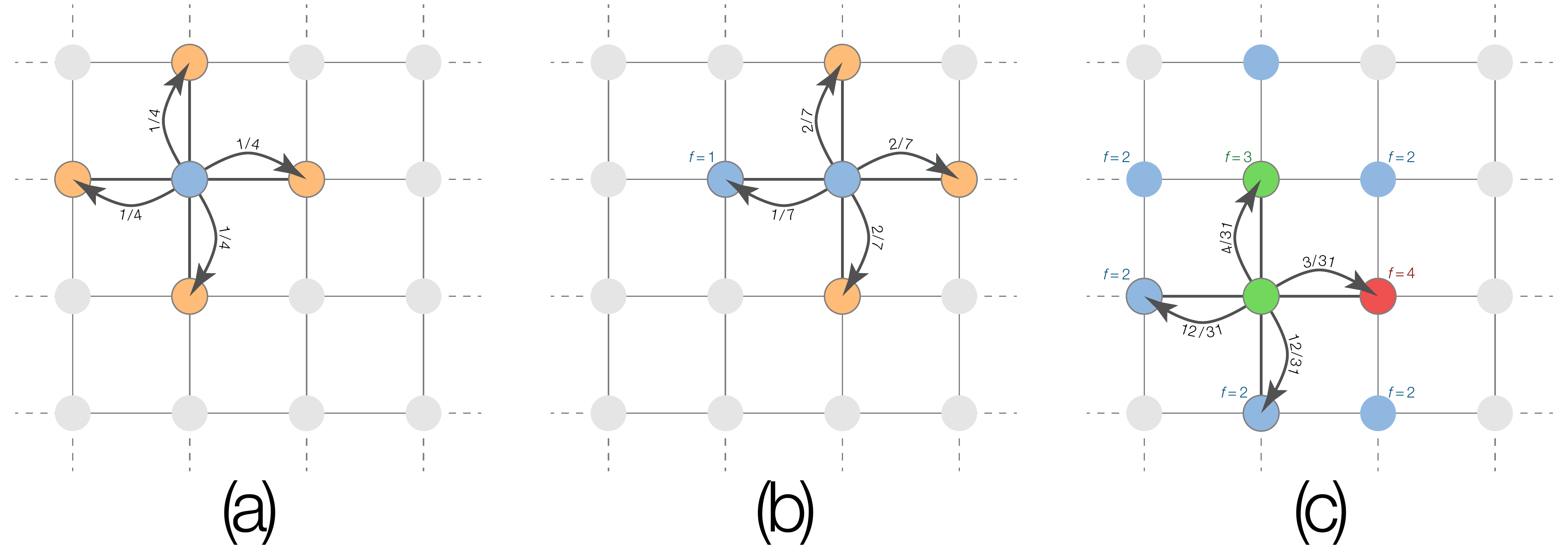}
 \caption{Example of the TSAW dynamics on a 2D lattice along three distinct time steps: (a) first iteration, (b) second iteration and (c) 13th iteration. Node colors indicate the number of times $f$ that a vertex was visited. The colors orange, blue, green, and red represent  $f=0$, $f=1$, $f=2$, and $f=3$, respectively. For each situation the permitted transitions are shown as arrows between nodes alongside with the transition probability given by equation~\ref{eq:tsawprobability}.}
 \label{fig:fig_trueSelfAvoiding}
\end{figure*}

In this study we consider the organization of the knowledge as a complex network, and simulate the researchers as agents under a TSAW dynamics. An example of the proposed TSAW dynamics for an agent is illustrated in Figure~\ref{fig:fig_trueSelfAvoiding}, where three situations of an agent under such dynamics are shown. Such example illustrates an agent and its respective transitions in the network, where we show how the transition probabilities modifications along some iterations.

In addition to the TSAW mechanism, we also incorporate the concept of stochastic flight~\cite{mandelbrot1983fractal,volchenkov2011random} to the proposed dynamics. Such a modification is needed to account for a scientific environment of multiple interacting agents, which  is usually done by reading each other papers, by attending conferences, or via collaborations~\cite{PhysRevE.88.012814,doi:10.1142/S0129183107010437,Viana2013371,0295-5075-99-4-48002,Newman16012001}. In particular, the flight dynamics is employed to simulate the events when a agent changes its object of study based on the discoveries made by another agent. In summary, the TSAW accounts for the local exploration of the knowledge network by a agent, while the stochastic flight allows agents to reach nodes farther away through jumps to knowledge entities close to the recent discoveries.

A variety of flight dynamics exists in the literature, including the L\'evy, Cauchy and Rayleigh flights~\cite{mandelbrot1983fractal}. In our model, we use a flight mechanism with a probability $\gamma$. If this mechanism is activated, the agent does not perform the TSAW dynamics at the current time step. As an alternative, the agent  jumps to an arbitrary destination node in the network.
The destination node is stochastically determined according to an influence field emanating from the other agents. In our model, each agent $v_j$ \emph{emits} a field $E_a(i)$ along the topological structure of the network to simulate knowledge dissemination. Such a field is defined in a way so that dissemination decays exponentially with the topological distance $d(i,j)$ from an observation vertex $v_i$ to $v_j$. Thus, for each agent, $a$, this field is computed as
\begin{equation} \label{eq:fielddef}
	E_a(i) = \sum_{j \in V}{\eta_i \exp(-d(i,j)  \tau) },
\end{equation}
with $V$ being the set of nodes in the knowledge network and $\tau$ a constant of the dynamics. The set of values $\eta_i$ constitutes another parameter of the dynamics and models the fitness of each agent, which allows us to simulate different fields magnitudes emanating from each agent. In other words, by considering this parameter, it is possible to emulate a case where the influence of the agents is not homogeneous. Here, the reduction of the influence field was modeled in terms of exponential decay instead of a power law, so as to constrain the spreading of the influence irrespectively to the network dimension~\cite{jackson1975electrodynamics,silva2012local,wei2014new}.

\begin{figure}[!htbp]
 \centering
 \includegraphics[width=1.\linewidth]{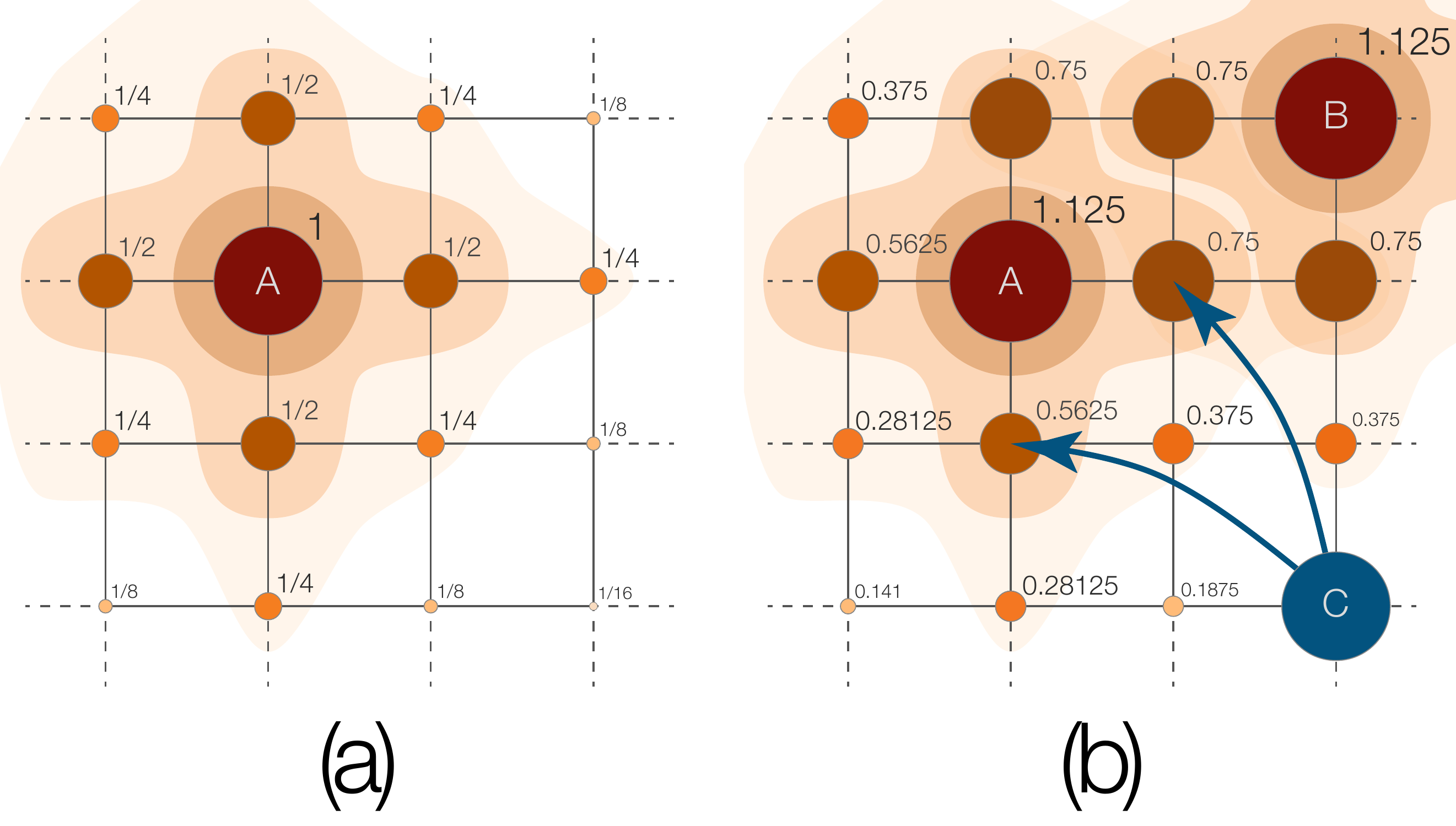}
 \caption{Example of fields emitted by agents. Item (a) represents the field emitted by a single agent, $E_a$. Note that for a given agent, the further the node,  weaker its respective field will be. Item (b) shows an example of a resultant field, $E$, which is a superposition of the fields generated by agents $A$ and $B$. Item (b) also illustrates agent $C$ being attracted by the resulting influence field. In this example $\eta_C = 0$ so agent $C$ has no influence over the others.}.
 \label{fig:fields}
\end{figure}

In the proposed model, the field acts as a stochastic attractor on the agents, inducing them to make a jump. The parameters $\tau$ and $\eta_i$ appearing in the definition of the field (equation~\ref{eq:fielddef}) can be interpreted, respectively, as the \emph{dissemination decay} and the individual \emph{fitness} of each agent.  The parameter $\tau$ defines the overall locality of the influence field. On the other hand, $\eta$ allows distinct performance to be assigned to each agent. The adopted influence field is illustrated in Figure~\ref{fig:fields}~(a).

For the simulations, we adopted a binary distribution of $\eta$, in which $\eta$ can assume only two possible values: $\eta_\text{common}$ and $\eta_\text{influential}$. A parameter $D_\eta$ defines the portion of agents having $\eta = \eta_\text{influential}$.

In our model, the fields are combined via superposition, thus, the total field $E(i)$ acting on a vertex $v_i$ can be calculated as
\begin{equation}
 E(i) = \sum_{a \in \mathcal{A}}{E_a(i)}.
\end{equation}
where $\mathcal{A}$ is the set of agents. An example of this field can be seen in Figure~\ref{fig:fields}~(b). Note that when the agent jumps it is attracted to the resultant field $E$ and tends to go to a position near the other agents. Finally, the destination of the jump is chosen among all the nodes of the network according to a probability $\Pi(i)$:
\begin{equation}
 \Pi(i) = \frac{E(i)}{\sum_{j \in V}{E(j)}}.
\end{equation}

In summary, our model starts with a set of $|A|$ agents randomly displaced over a network representing the knowledge structure. For each iteration, the agents walk along the network according to the two proposed dynamics, TSAW or flight, depending on the jump probability $\gamma$. We should emphasize that the TSAW dynamics in our model is not dependent on the interaction between agents. Such influence takes place only during the decision process of a jump.

\subsection{Network database}
In order to characterize the performance of our model among different knowledge organizations, we put together a small dataset of networks having many distinct properties. This selection starts with one of the simplest network models, a bidimensional lattice (LA), in which nodes are regularly distributed over a squared grid and are connected by proximity. In this work, each node in such a network is connected to its nearest four nodes, except those lying at the borders of the networks. In addition, in order to maintain the same degree in the whole network and to eliminate the boundary effect, we also incorporate a lattice to our dataset, with the border nodes being toroidally connected, which is henceforth called toroidal Lattice (TLA). In this case, all nodes of the network become indistinguible among themselves. We also incorporate some random network models to the dataset, namely the Watts-Strogatz (\emph{WS})~\cite{watts1998WS}, the Barab\'asi-Albert model (\emph{BA})~\cite{barabasi1999BA}, the community model developed by Lancichinetti \emph{et al} (\emph{CN})~\cite{silva2016using} and the Waxman model (\emph{WAX})~\cite{Waxman1988Connections}.

The WS model reproduces the small-world phenomenon commonly found in many real-world networks, e.g. social network~\cite{wasserman1994social}, food web~\cite{dunne2002food}, brain networks~\cite{sporns2004small,bassett2006small}, among others~\cite{amaral2000classes}. In such networks, most of the nodes can be reached from the other nodes in a small number of steps~\cite{costa2007characterization}.

Another important characteristic present in many real-world networks is their scale-free nature. The BA model reproduces this characteristic through two mechanisms: preferential attachment and network growth. More specifically, new nodes are progressively added to the network and are more likely to connect with nodes already presenting many connections, resulting in a power law node degree distribution~\cite{costa2007characterization}.

We also incorporated a geographic model to the dataset. In this case, we choose the traditional Waxman model~\cite{Waxman1988Connections}, in which nodes are randomly displaced along a two dimensional space and are connected according to a probability that decays exponentially with the distance between each pair of nodes.

In addition to the traditional complex network models, we also considered the use of networks presenting community structure, which can be found in many real-world networks representing knowledge, such as citation networks~\cite{silva2016using} and Wikipedia~\cite{silva2011investigating}. In order to do so, we generated benchmark networks of Lancichinetti~et~al.~\cite{lancichinetti2008benchmark}, as we call Community Networks (CN). This model generates scale-free networks with a given number of communities. In addition, there are other parameters in this model, such as the mixing parameter, $\mu$, which defines how often nodes from a community connect with nodes from other communities. Furthermore, there are parameters to control the average degree of the network, $k_{min}$ and $k_{max}$ which are minimum and maximum degree, respectively, and the minimum and maximum community size, $s_{min}$ and $s_{max}$, respectively, where the condition $s_{min} > k_{min}$ and $s_{max} > k_{max}$ should be enforced. Further information about this model can be found in~\cite{lancichinetti2008benchmark}.

We also included in our dataset two real-world networks representing knowledge: a Wikipedia~\cite{silva2011investigating} (\emph{WIKI}) network and a citation network obtained from the Web of Science~\cite{silva2016using} (\emph{WOS}). The WIKI network was obtained from a subset of the Wikipedia incorporating only articles from two main categories:  Biology and Mathematics. In such a network, each node represents an article and the edges indicate a reference between two articles. The WOS network was obtained from the set of articles and citations resulting from a query on the Web of Science encompassing only the complex networks field.

\begin{figure*}[!htbp]
 \centering
 \includegraphics[width=0.80\linewidth]{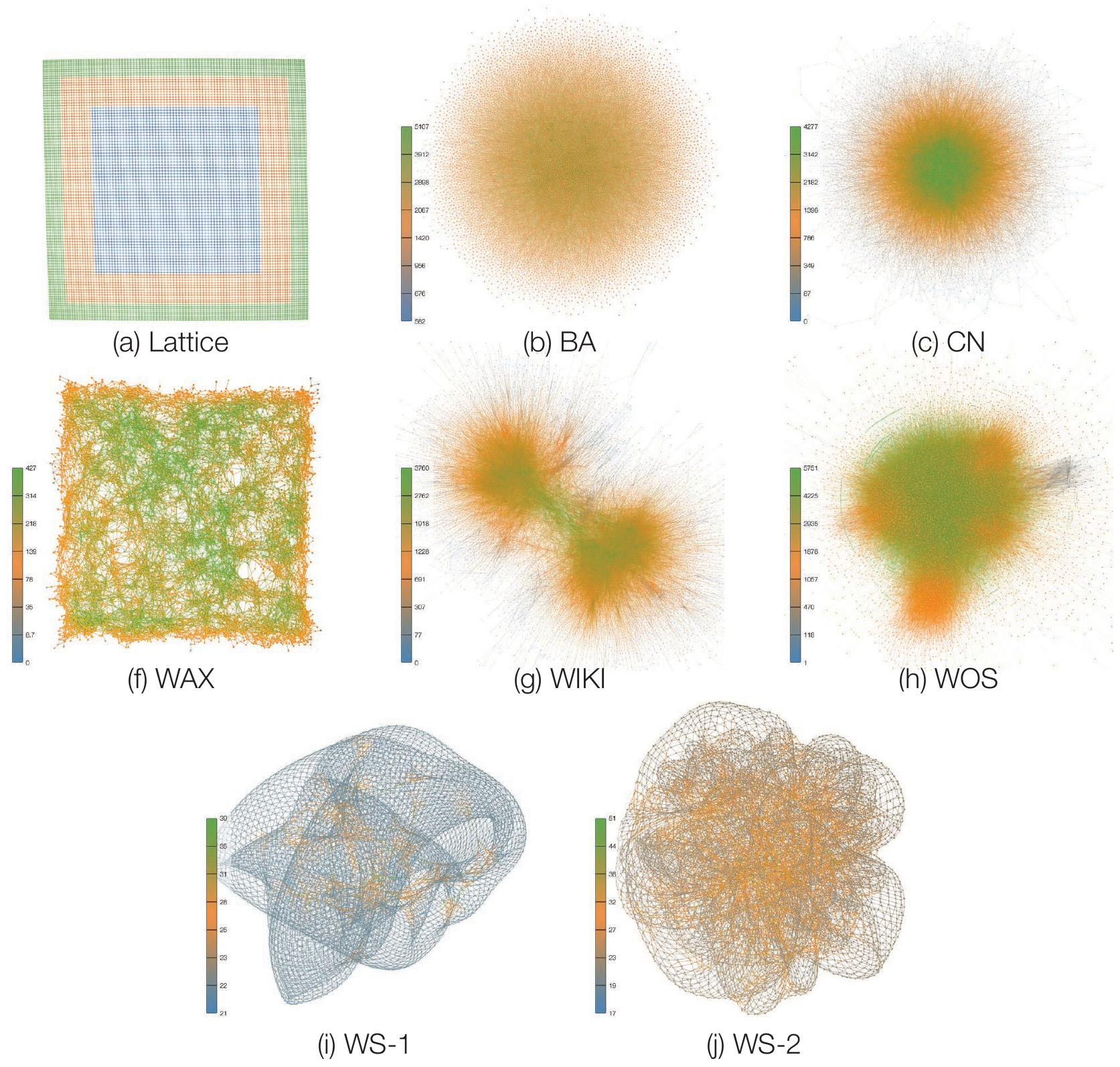}
 \caption{Visualization of all selected networks, with colors representing the accessibility of each node ($A_i^{(h=3)}$).}
 \label{fig:visualizations}
\end{figure*}

The employed models were configured to have similar number of nodes as in the selected real networks. In addition, we considered all networks as unweighted and undirected. The dataset of networks is summarized in table~\ref{tab:networks}. Additionally, we also obtained a visualization for each network, which is shown in Figure~\ref{fig:visualizations}.

\begin{table*}[]
\centering
\caption{Description of the analyzed networks. The number of nodes ($N$) and the average network degree ($k$) are shown. }
\label{tab:networks}
\begin{tabular}{c p{7.0cm} c c c c}
\hline
Network & Description                                                                      & $N$ & $k$ & Parameters & Refs. \\ \hline
LA      & Bidimensional lattice                                                            & 10k & 3.96  &            &   -- \\
TLA     & Bidimensional toroidal lattice                                               & 10k & 4.00  &            &   --   \\
WS-1    & Watts-Strogatz model                                                       &   10k  &  4.00  &   $p=0.001$         &    \cite{watts1998WS}   \\
WS-2    & Watts-Strogatz model                                                        &   10k  &  4.00   &  $p=0.005$     &  \cite{watts1998WS}     \\
WAX     & Waxman model                                                                  &  $\approx$10k  &   6.02  &     $\alpha = 1$, $\beta = 0.015$, and $L=1$       &   \cite{Waxman1988Connections}    \\
BA        & Barabasi-Albert model                                                       &  10k   &  6.00 &            &  \cite{barabasi1999BA}     \\
CN    & Fortunato model                                                                 &   10k  &  5.63   &    2 communities,  $\mu = 0.2$       &   \cite{lancichinetti2008benchmark}    \\
%CN-2    & Fortunato model                                                                 &   10k  &  5.69  &    3 communities,  $\mu = 0.2$       &   \cite{lancichinetti2008benchmark}    \\
%CN-3    & Fortunato model                                                                 &  10k   &  5.61  &    14 communities, $\mu = 0.2$       &   \cite{lancichinetti2008benchmark}    \\
WIKI    & Subset of wikipedia encompassing articles from Biology and Mathematics               &  $\approx$12k   &  7.29   &      --     &    \cite{silva2011investigating}  \\
WOS     & Citation network obtained for the query "Complex Networks" on the Web of Science &  $\approx$11k   & 17.08    &     --       &  \cite{silva2016using}     \\
\hline
\end{tabular}
\end{table*}

\subsection{Performance evaluation}
In order to characterize the performance of the proposed dynamics for distinct parameters configurations, we employed a measurement to quantify the learning performance of the agents. The dynamics is evaluated according to their collective discovery performance, more specifically the total number of explored nodes, $\varepsilon_T$, which is defined as
\begin{equation}
 \varepsilon_T(t_{a}) = \sum_{t=1}^{t_{a}}{\varepsilon(t)},
\end{equation}
where $t_{a}$ is the actual time (i.e. the current iteration) and $\varepsilon(t)$ is the number of newly explored nodes at $t$, not taking into account those explored on iterations before $t$.

A full understanding of the presented dynamics can be accomplished by studying how the learning performance is affected by distinct topological characteristics, such as by how central a node is in a network, as the dynamics may vary substantially depending on the region of the network they are being analyzed. Complex networks can display very distinct local characteristics that may affect the dynamics arising from their structure. For instance, it is known that a random walk dynamics can be employed to find the community structure of a given network~\cite{rosvall2008maps}. For this reason, we also considered the evaluation of the collective discovery for distinct regions of the networks. In this case, each region corresponds to sets of nodes sharing a similar property, which can be quantified by a topological measurement. In this work, we use the accessibility~\cite{rodrigues2009structure} measurement to characterize distinct regions of the networks. Such measurement is known to detect borders and influential nodes in a high variety of real-world networks~\cite{travenccolo2009border,de2014role,1742-5468-2015-3-P03005,travenccolo2008accessibility}. In our study we estimate the accessibility as in~\cite{travenccolo2008accessibility}. The accessibility of a vertex $i$ is defined as
\begin{equation}
 A^{(h)}_i = \exp \left(- \sum_j p^{(h)}_{ij} \log p^{(h)}_{ij} \right),
\end{equation}
where $p^{(h)}_{ij}$ is the probability of reaching a vertex $j$ having departed from $i$, after $h$ steps. In this study, we considered $h=3$ to avoid the limited size effects as the diameter of some networks in our database can reach very low values, such as $6$ or even $5$.

\section{Results and discussion}
We compared the learning performance of our dynamics among different networks and for many distinct sets of parameters. Moreover, by considering the structure of each network, we discuss the performance of the proposed dynamics in different network regions. Henceforth, the analyses take into consideration $300$ realizations for each combination of parameters.   As a result, the performance of the dynamics is measured in terms of the average total number of explored nodes ($\langle\varepsilon_T\rangle$) calculated over all these realizations.

\subsection{Network dynamics evaluation}
\begin{figure}[!htbp]
 \centering
 \subfigure[$D_\eta$.]{\includegraphics[width=0.65 \linewidth]{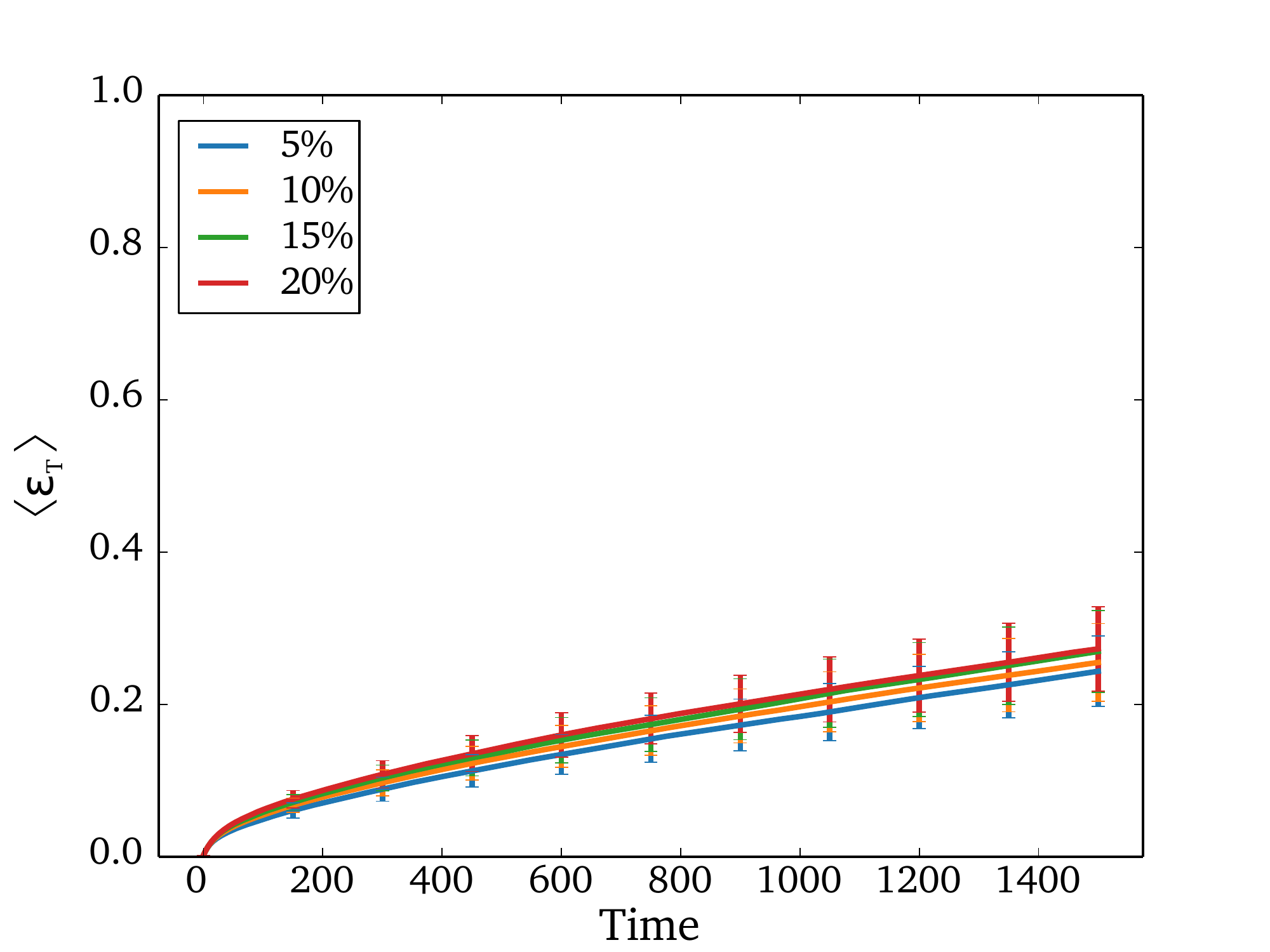}}
 \subfigure[$\tau$.]{\includegraphics[width=0.65 \linewidth]{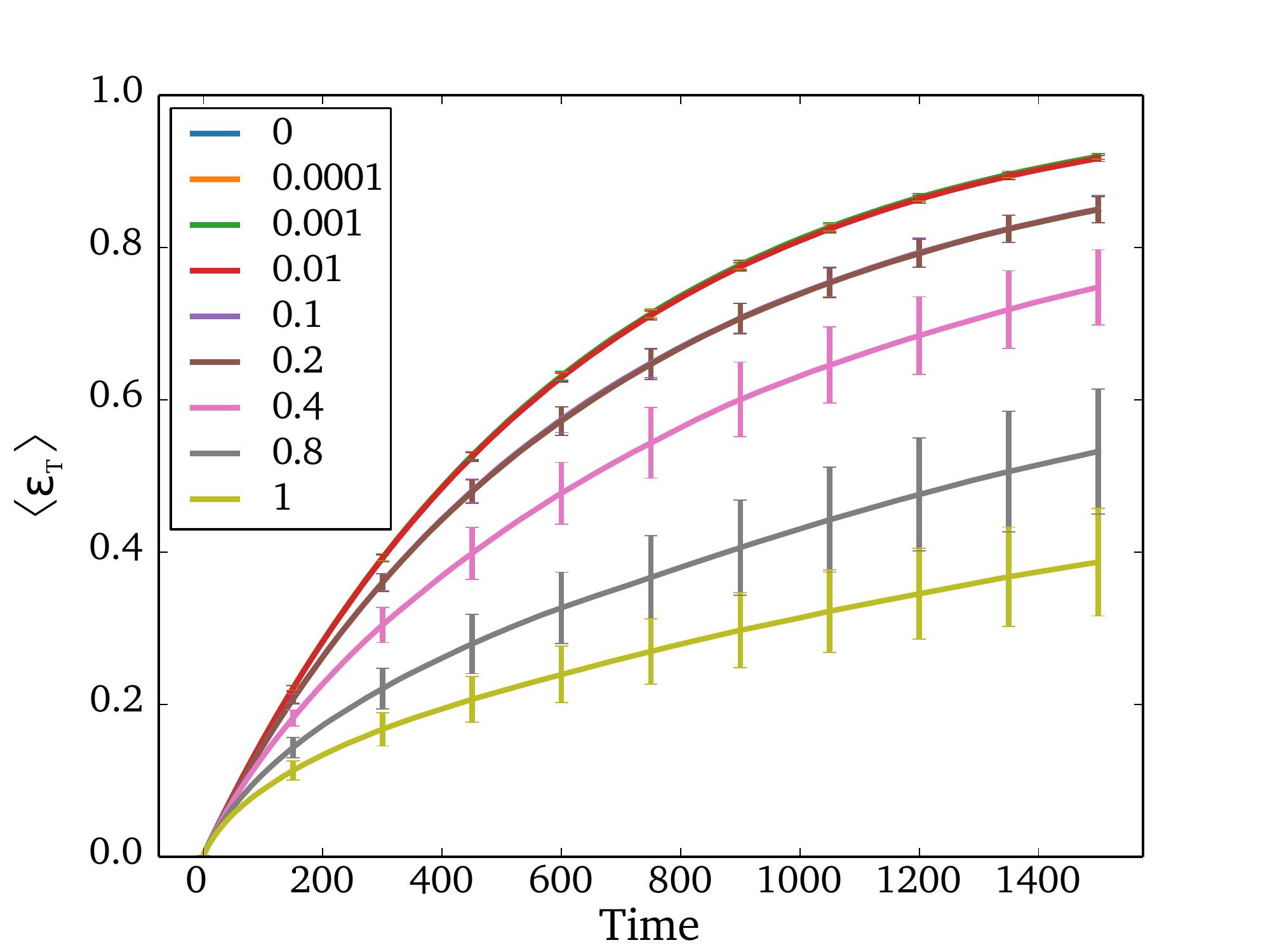}}
 \caption{Representations of the dynamics executed in LA, where the items (a) and (b) represent the results of the parameter changes of $D_\eta$ and $\tau$, respectively.}
 \label{fig:fig_lattice}
\end{figure}

\begin{figure*}[!htbp]
 \centering
 \subfigure[LA]{\includegraphics[width=0.32\linewidth]{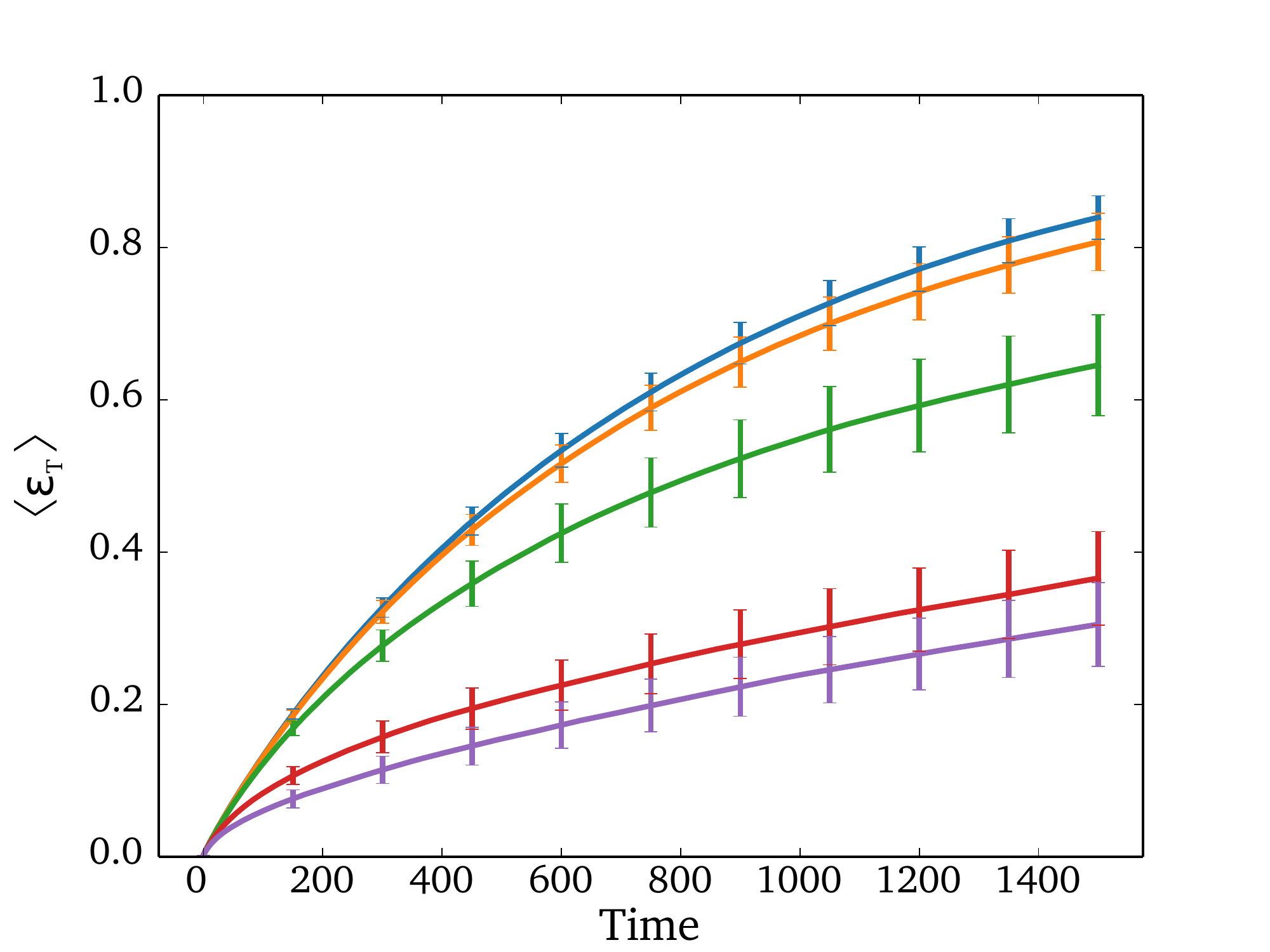}}
 \subfigure[WS-1]{\includegraphics[width=0.32\linewidth]{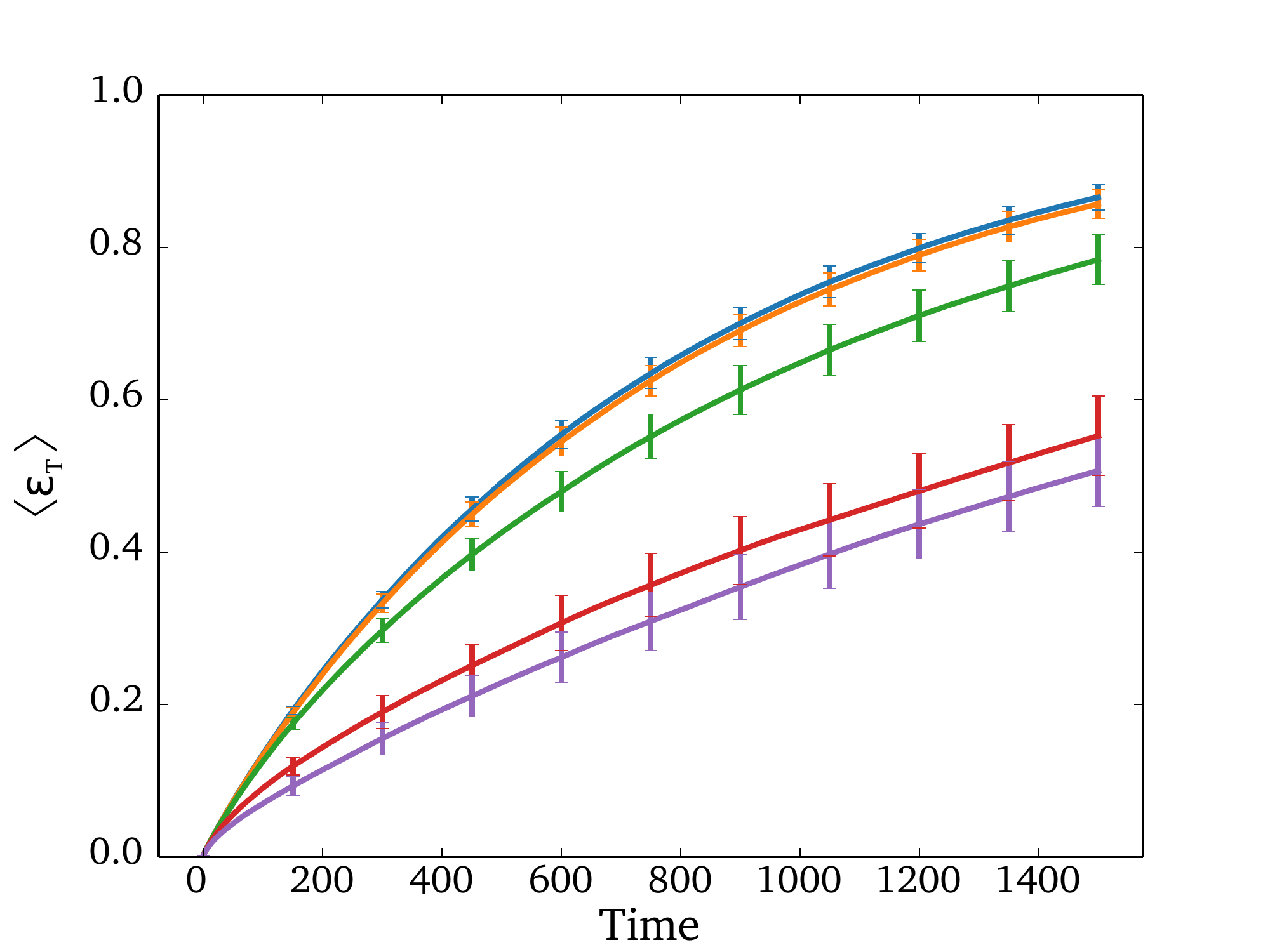}}
 \subfigure[WS-2]{\includegraphics[width=0.32\linewidth]{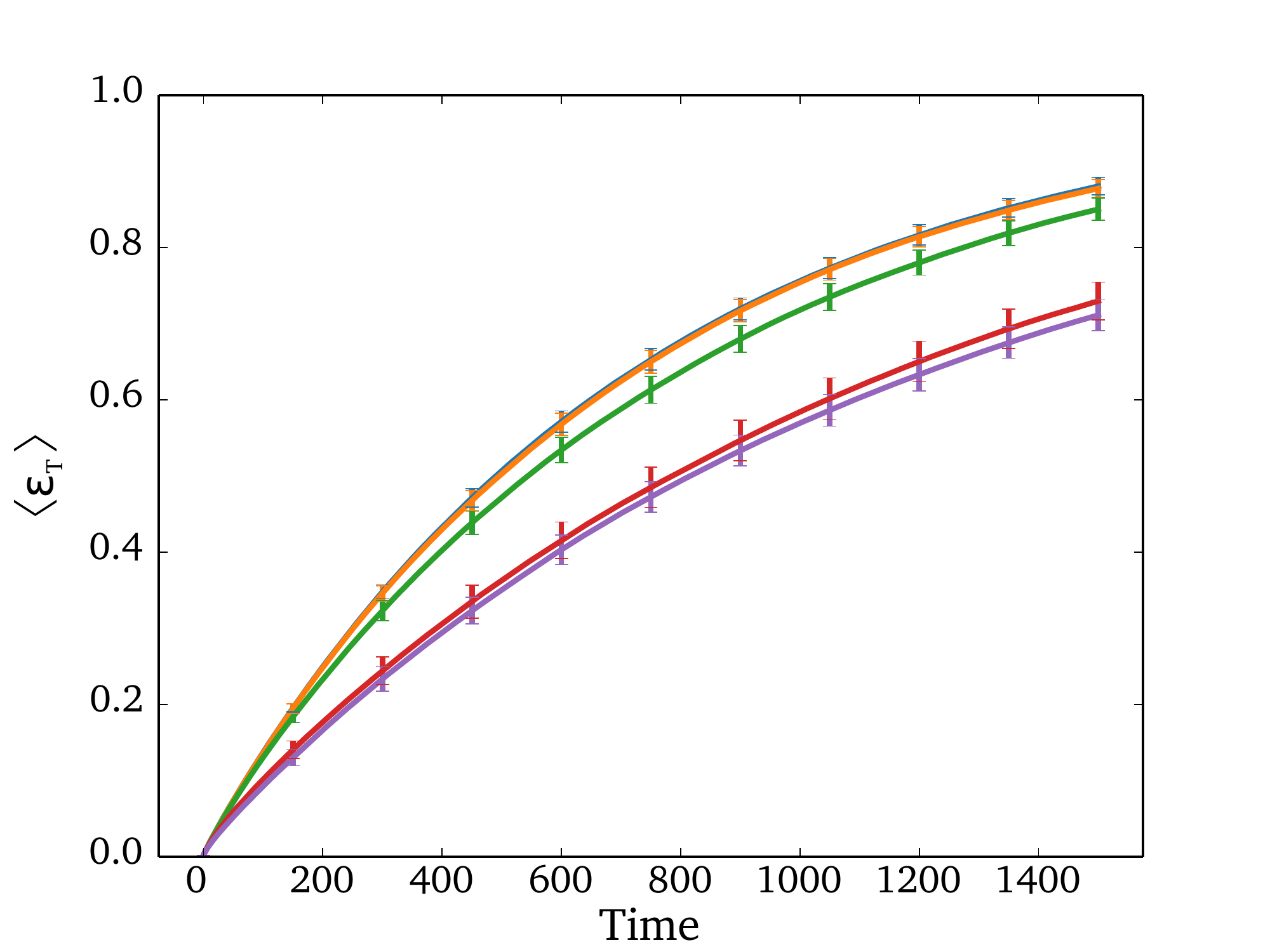}}
 \subfigure[WOS]{\includegraphics[width=0.32\linewidth]{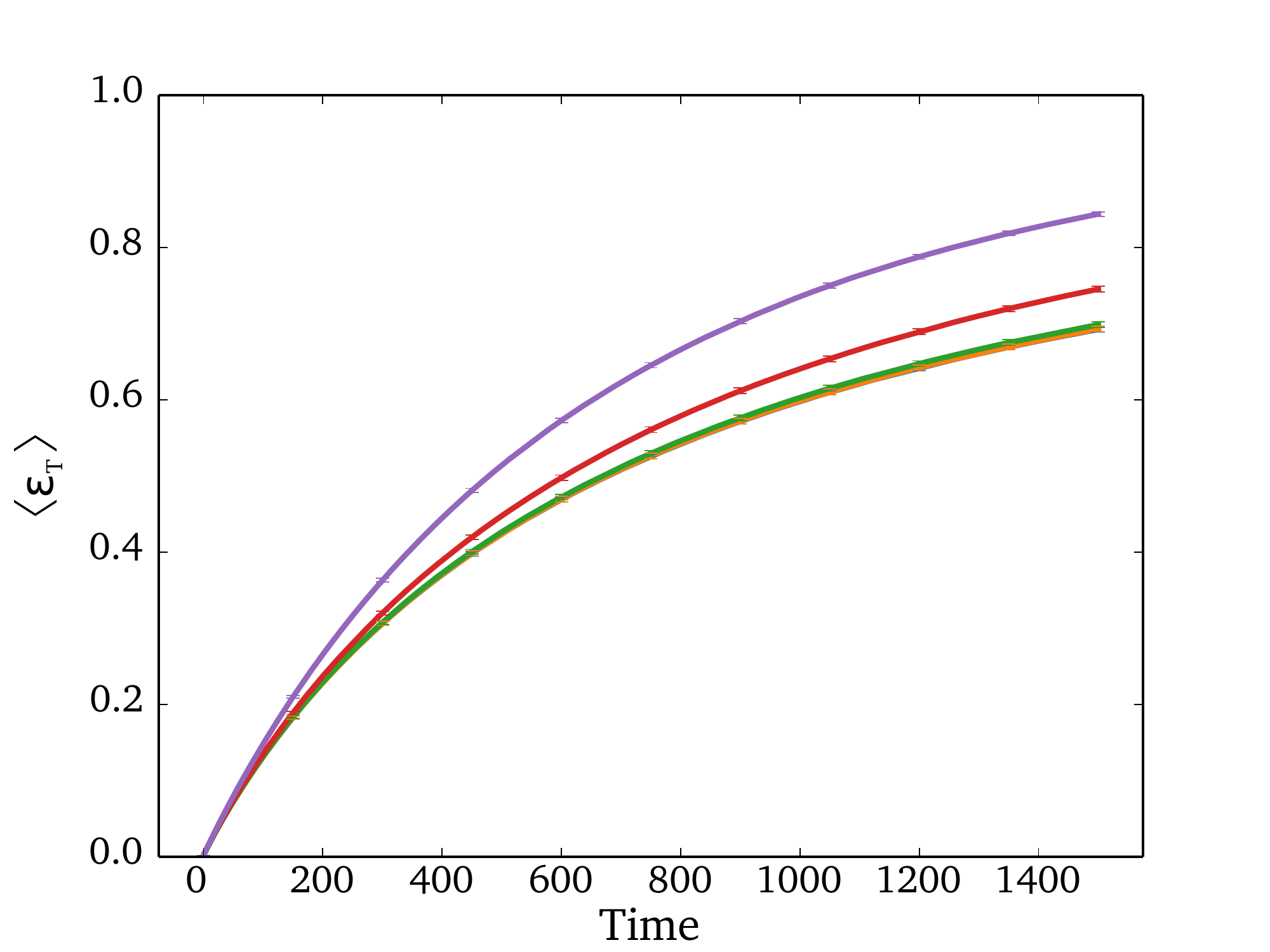}}
 \subfigure[WIKI]{\includegraphics[width=0.32\linewidth]{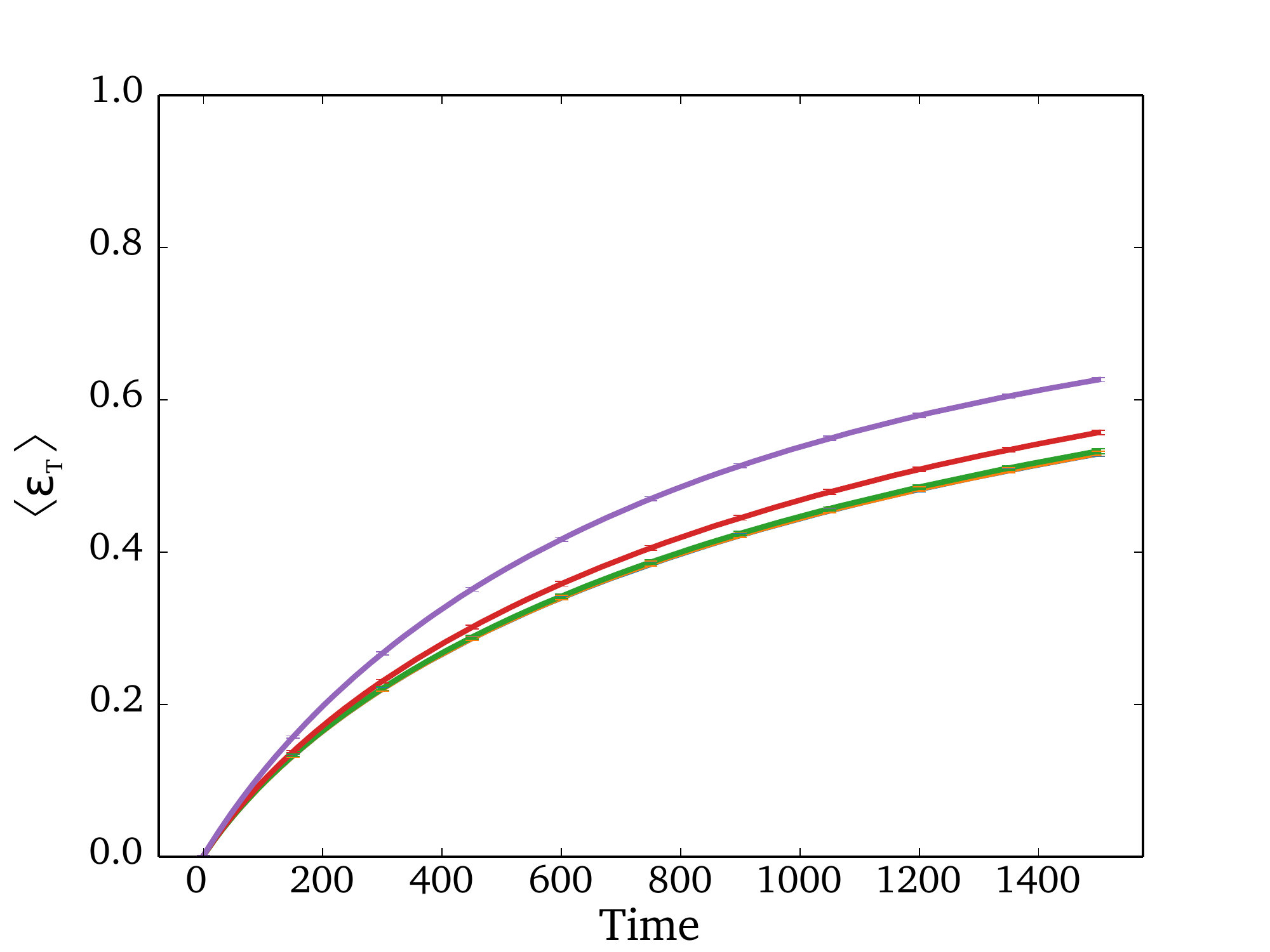}}
 \subfigure[CN]{\includegraphics[width=0.32\linewidth]{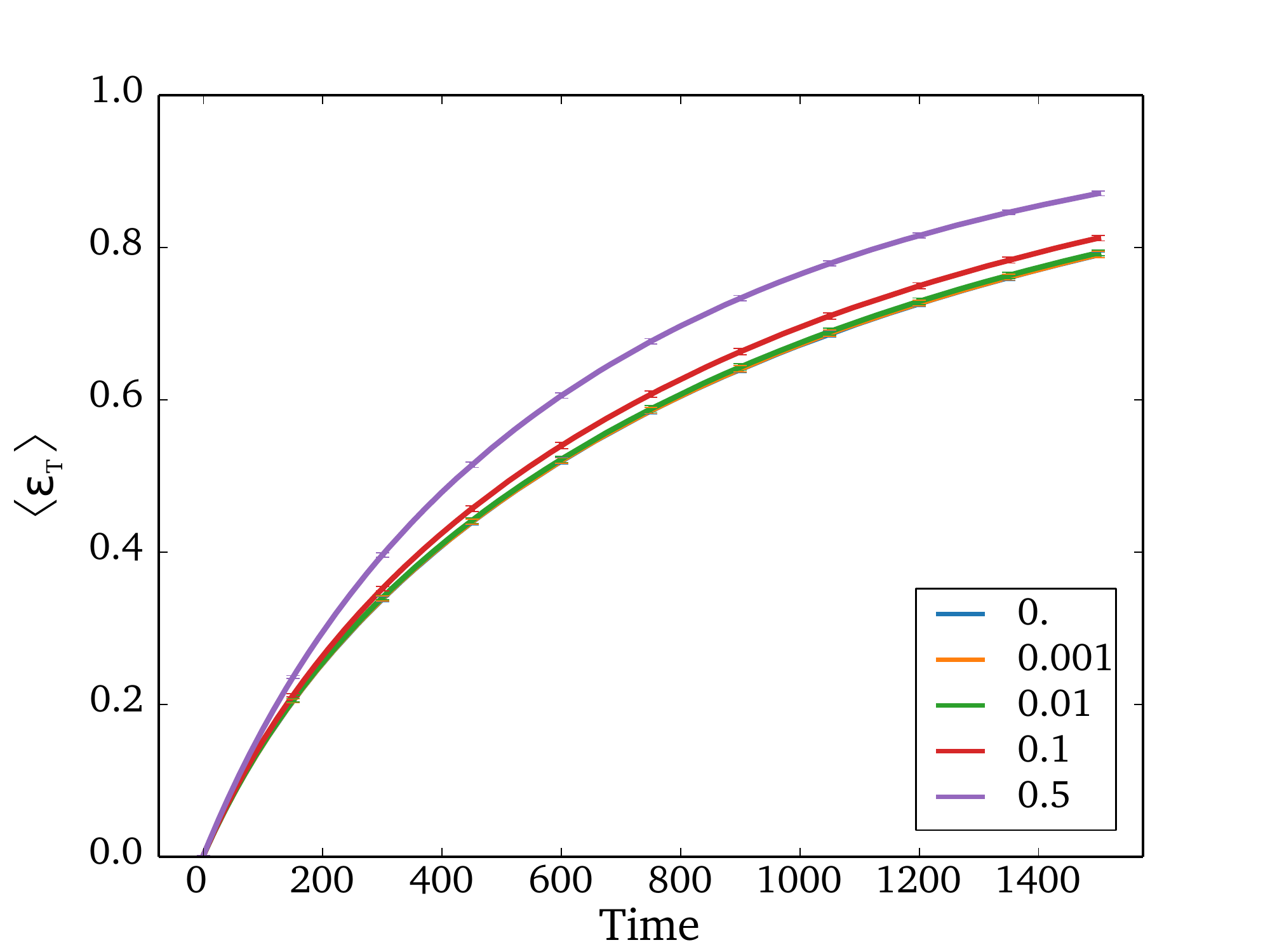}} \caption{Average performances of the proposed dynamics ($\langle\varepsilon_T\rangle$) and corresponding standard deviations obtained by varying the jump probability ($\gamma$) considering six networks: (a)LA; (b)WS-1, (c)WS-2, (d)WOS, (e)WIKI and (f)CN. The parameter $\gamma$ is indicated by colors according to the legend.}
 \label{fig:fig_geo_jump}
\end{figure*}

We first analyzed the lattice network because of its simplicity.  We found that the knowledge acquisition performance $\langle\varepsilon_T\rangle$ in LA and TLA did not vary given several combinations of parameters ($D_\eta$, $\tau$, and $\gamma$). For this reason, we only present the results obtained for the LA network. First, we observed that the variations of $D_\eta$ parameter were weakly reflected in $\langle\varepsilon_T\rangle$, as shown in Figure~\ref{fig:fig_lattice}~(a). We can infer that the variation of the number of agents with high influence, represented by the vector $D_\eta$, does not change the dynamics performance. The evaluated performance of the other parameters, $\tau$ (the locality of the field) and  $\gamma$ (the jump probability), are shown in Figure~\ref{fig:fig_lattice}~(b) and Figure~\ref{fig:fig_geo_jump}~(a), respectively. Even though all the curves present high standard deviation, the dynamics performance,  $\langle\varepsilon_T\rangle$, was greatly affected by varying $\tau$ or $\gamma$. We also observe that the performance decreases when any of the these two parameters increases. Regarding the parameters $D_\eta$ and $\tau$, the results obtained for the spatial models, WS-1, WS-2 and WAX are markedly similar to those obtained by the analysis of the LA model.

\begin{figure}[!htbp]
 \centering
 \includegraphics[width=0.65\linewidth]{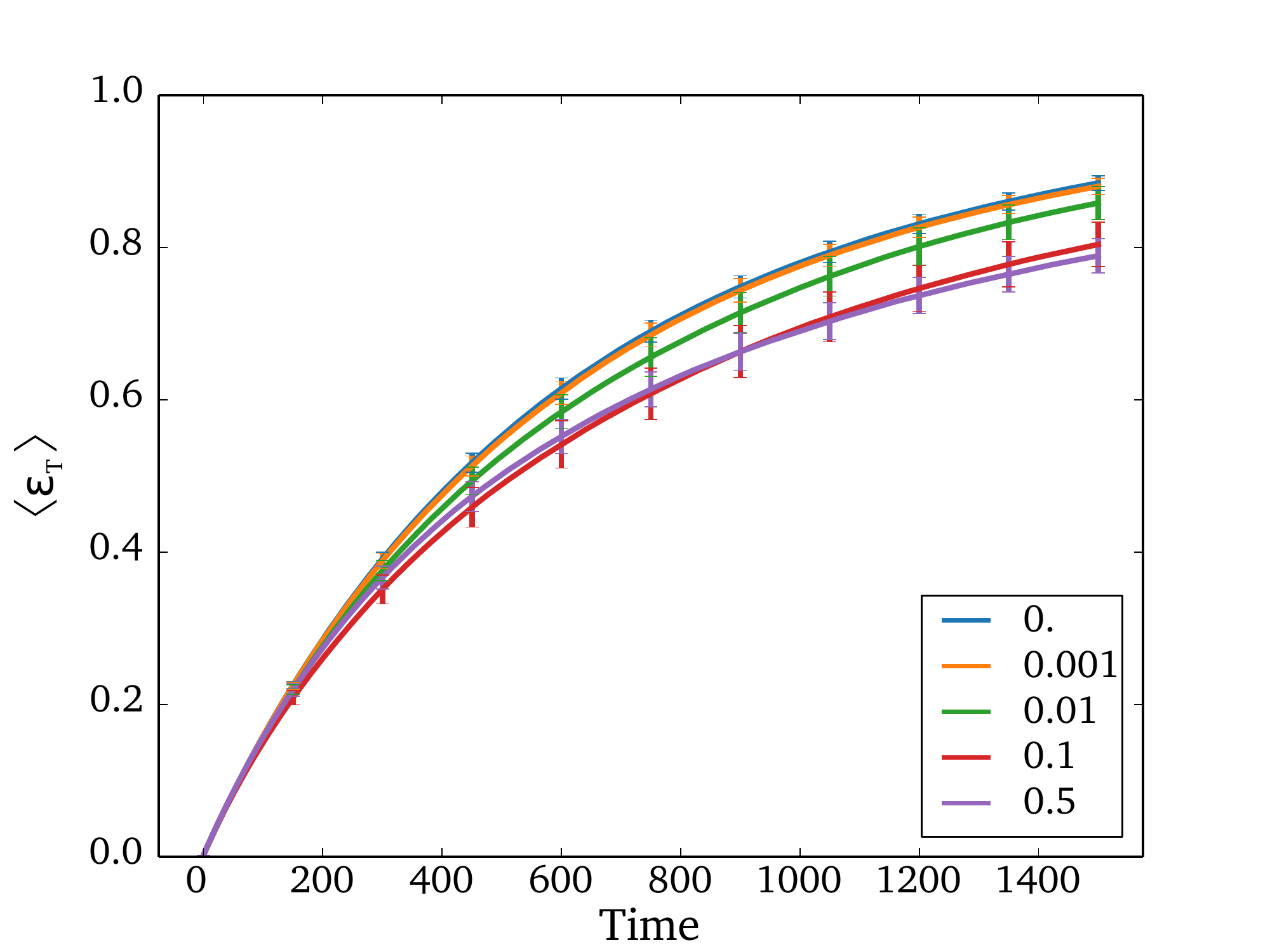}
 \caption{Performance, $\langle\varepsilon_T\rangle$, of our dynamics, varying the values of $\gamma$ adopted in the WAX networks.}
 \label{fig:fig_wax}
\end{figure}

By focusing the analysis on the $\gamma$ parameter, we found similarities among the curves of $\langle\varepsilon_T\rangle$ obtained for the networks LA, WS-1, and WS-2, as shown in Figure~\ref{fig:fig_geo_jump}(a-c). In these cases, we also observed that the standard deviations of $\langle\varepsilon_T\rangle$ are similar among all configurations. In addition, increasing $\gamma$ was found to improve the dynamics performance significantly for these networks. Despite these similarities, the impact of increasing $\gamma$ is diminished as the probability of rewiring increases for the WS model (recall that LA also corresponds to a WS model with zero rewiring probability). This indicates that $\langle\varepsilon_T\rangle$ tends to be more sensitive to the $\gamma$ parameter when the network is more spatial or organized. However, the results obtained for the WAX, shown in Figure~\ref{fig:fig_wax}, indicate that this behavior is not a direct consequence of the spatiality of networks. The WAX network, a spatial structure, presented particularly low performance variation with the increase of $\gamma$. As in the previous cases, it is better to explore the network without using jumps. Furthermore, we also observed that varying $\tau$ produced similar results.

\begin{figure*}[!htbp]
 \centering
 \subfigure[$\gamma$]{\includegraphics[width=0.32\linewidth]{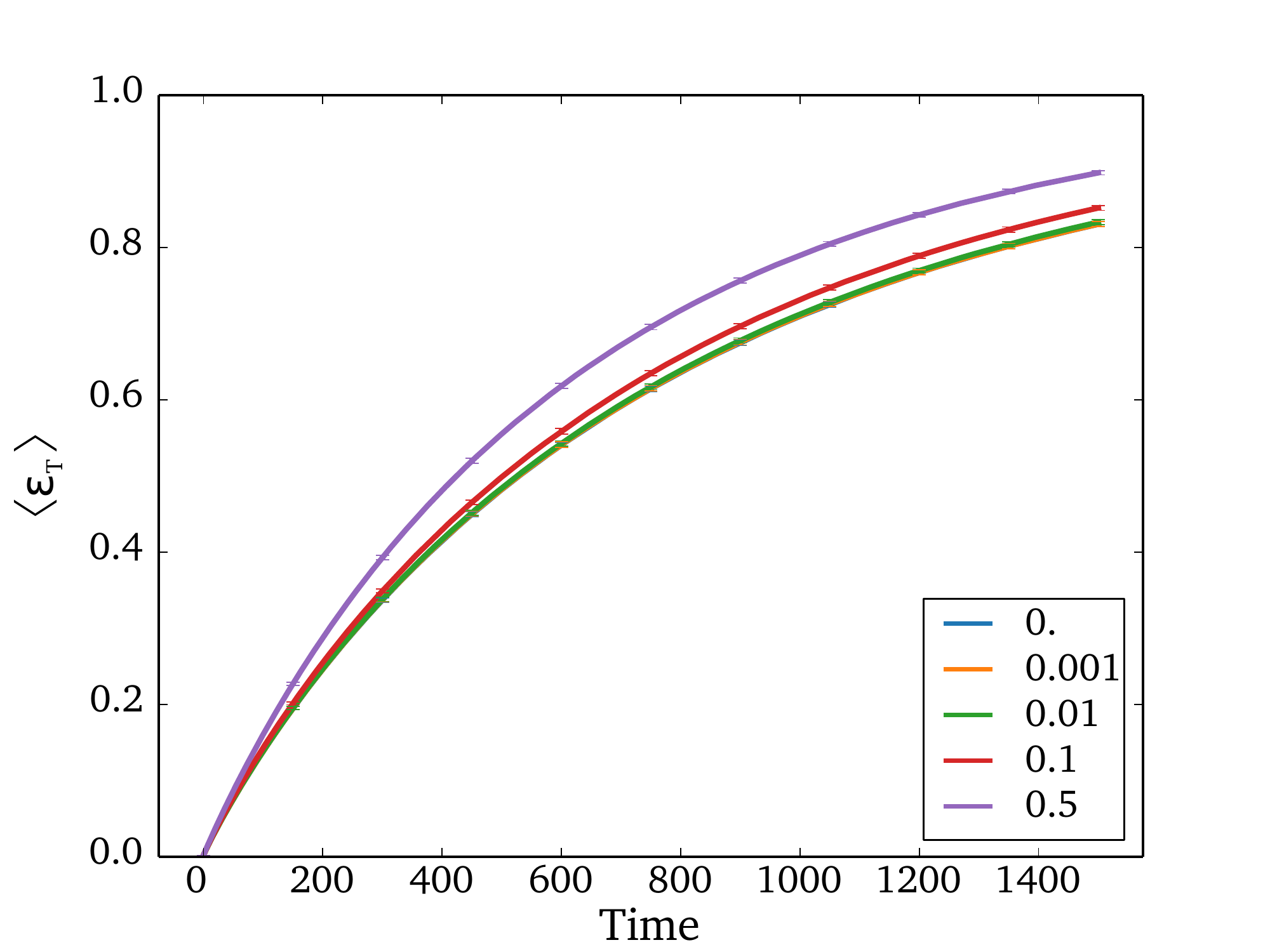}}
 \subfigure[$D_\eta$]{\includegraphics[width=0.32\linewidth]{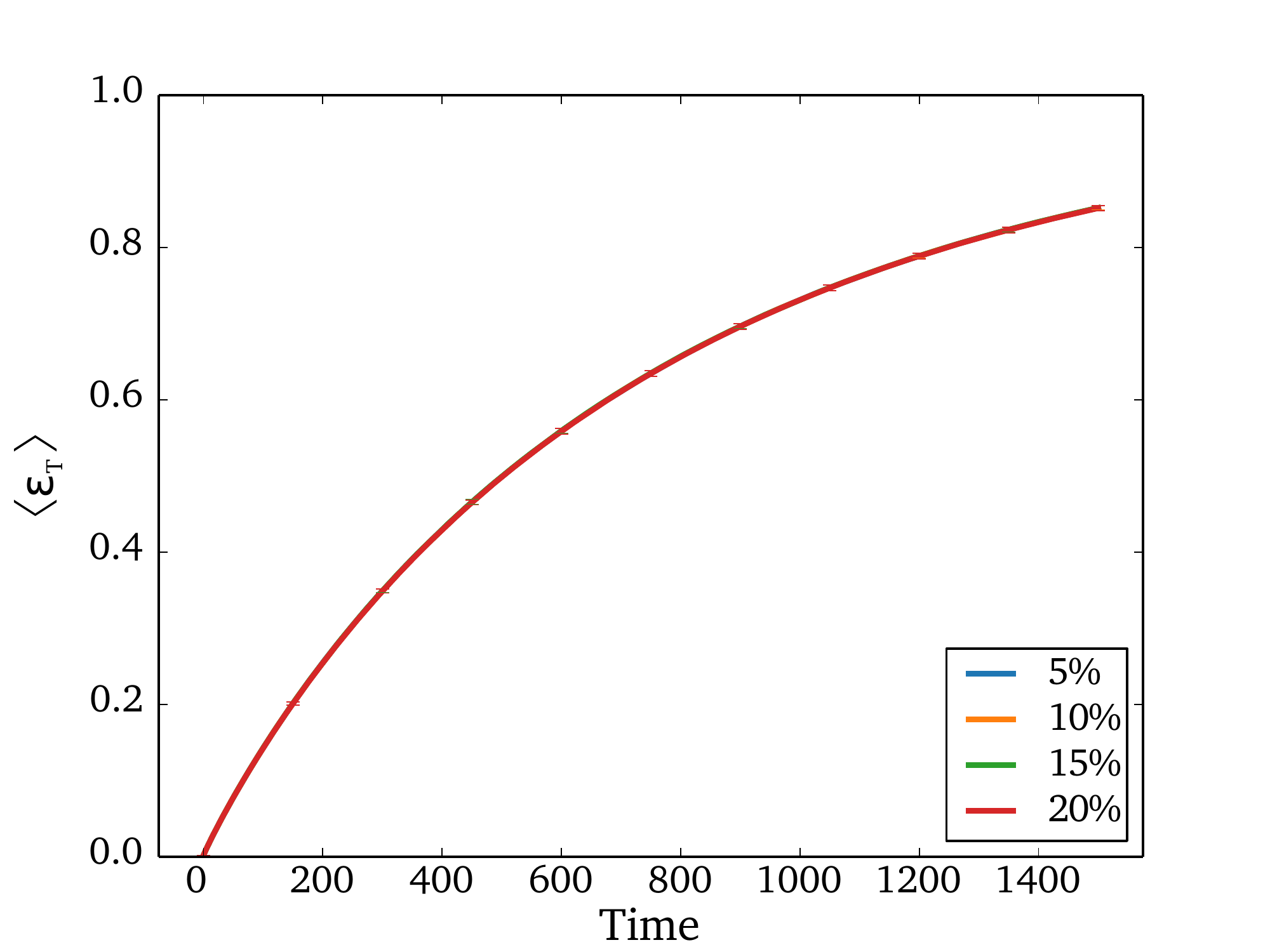}}
 \subfigure[$\tau$]{\includegraphics[width=0.32\linewidth]{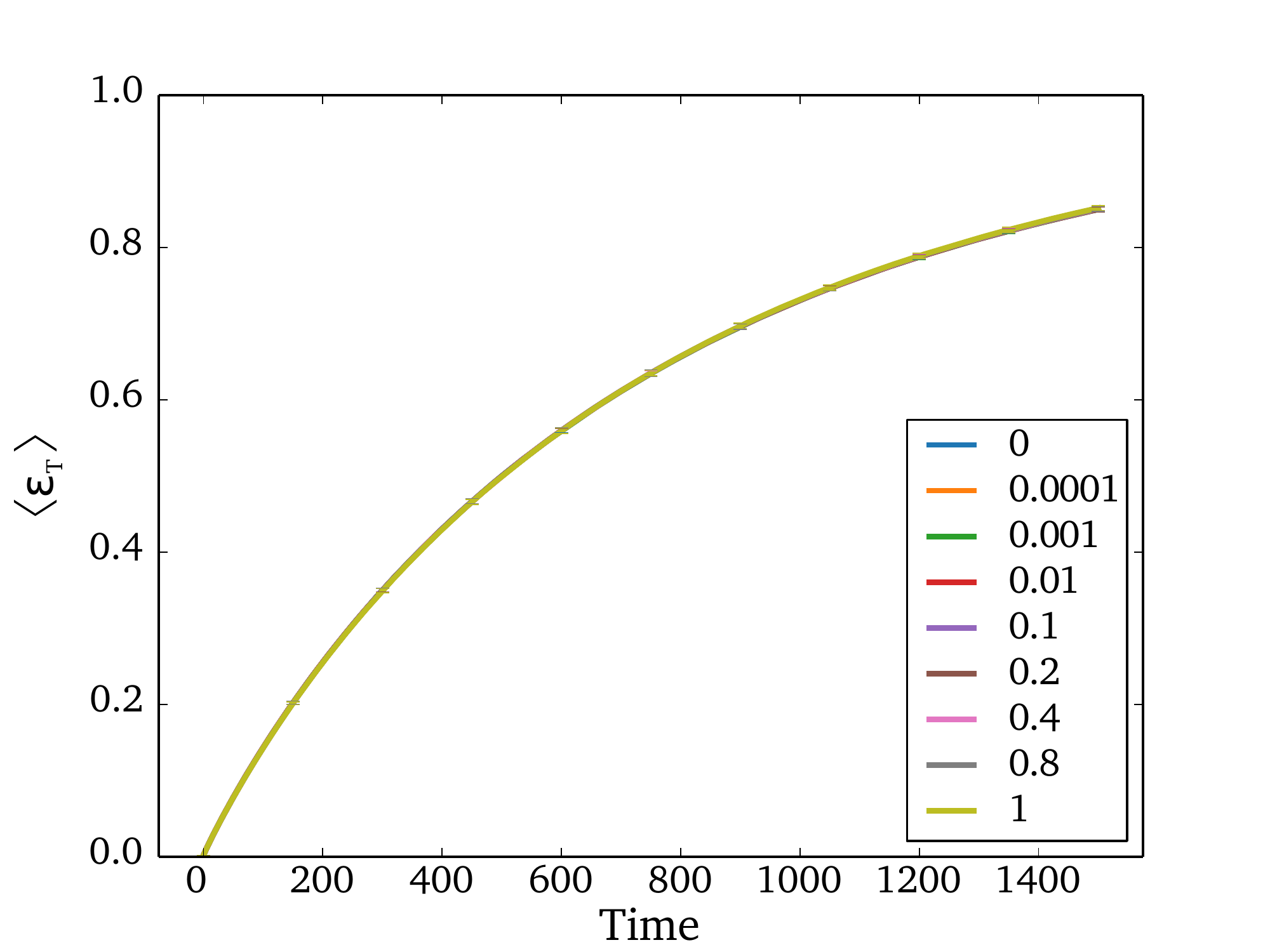}}
 \caption{The performance, $\langle\varepsilon_T\rangle$, obtained from the simulation of our dynamics on the BA network while varying the parameters: (a)$\gamma$,  (b)$D_\eta$ and (c)$\tau$. We note that the influence of the variations of $D_\eta$ and $\tau$ were not significative. On the other hand, the network exploration is more efficient when the jump probability $\gamma$ is high.}
 \label{fig:fig_ba}
\end{figure*}

We also analyzed how the performance can be affected by varying the jump probability ($\gamma$) for the WOS, WIKI and CN networks.  In contrast to the spatial networks discussed above, these networks are inherently more related to knowledge structures. The performance curves obtained for these networks are shown in Figure~\ref{fig:fig_geo_jump}(d-f). Differently from the results obtained for spatial networks, increasing $\gamma$ improves the performance.  We note, however, that the effects of changing $\gamma$ is much lower than for the spatial networks. The curves obtained for the BA network also present this same behavior, as shown in Figure~\ref{fig:fig_ba}(a). In addition, the WIKI network presented a general performance substantially lower than the other models even when compared with the CN network, which reproduces many of the WIKI network characteristics, such as the number of nodes, node degree distribution and its community structure.

Regarding the other parameters, we found no significant influence of $D_\eta$ and $\tau$ to the dynamics performance for the considered knowledge networks. In these cases, the curves are very similar to those obtained for the BA network, which are shown in Figures~\ref{fig:fig_ba}(b)~and~(c).

\subsection{Dynamics evaluation in network regions}
We analyzed above the networks in terms of the average performance of $\langle\varepsilon_T\rangle$ by measuring the dynamics performance globally.   Now, we consider different network regions. For this purpose, we understand a network region as being defined by a set of nodes presenting a certain topological characteristic (such as being at the borders of the network, or presenting similar accessibility). To measure the performance, we consider the value of $\langle\varepsilon_T\rangle$ after $1000$ iterations of the dynamics.  This number of iterations was chosen because, if we consider a small number of iterations, the exploration becomes local and, consequently, similar for all considered networks. On the other hand, for high values, the regions are totally explored resulting in similar performance.

Considering our database, we start the analysis with the LA model. The toroidal configuration is not taken into account because it is impossible to define particular regions in this network. For the LA model, we calculated the Chebyshev distance~\cite{prtools}, known as maximum value distance, between the position of each node and the geographical center of the network, instead of accessibility measurement. In order to do so, we used the average position of all network nodes as the geographical center.

\begin{figure}[!htbp]
 \centering
 \subfigure[$\gamma$]{\includegraphics[width=0.85\linewidth]{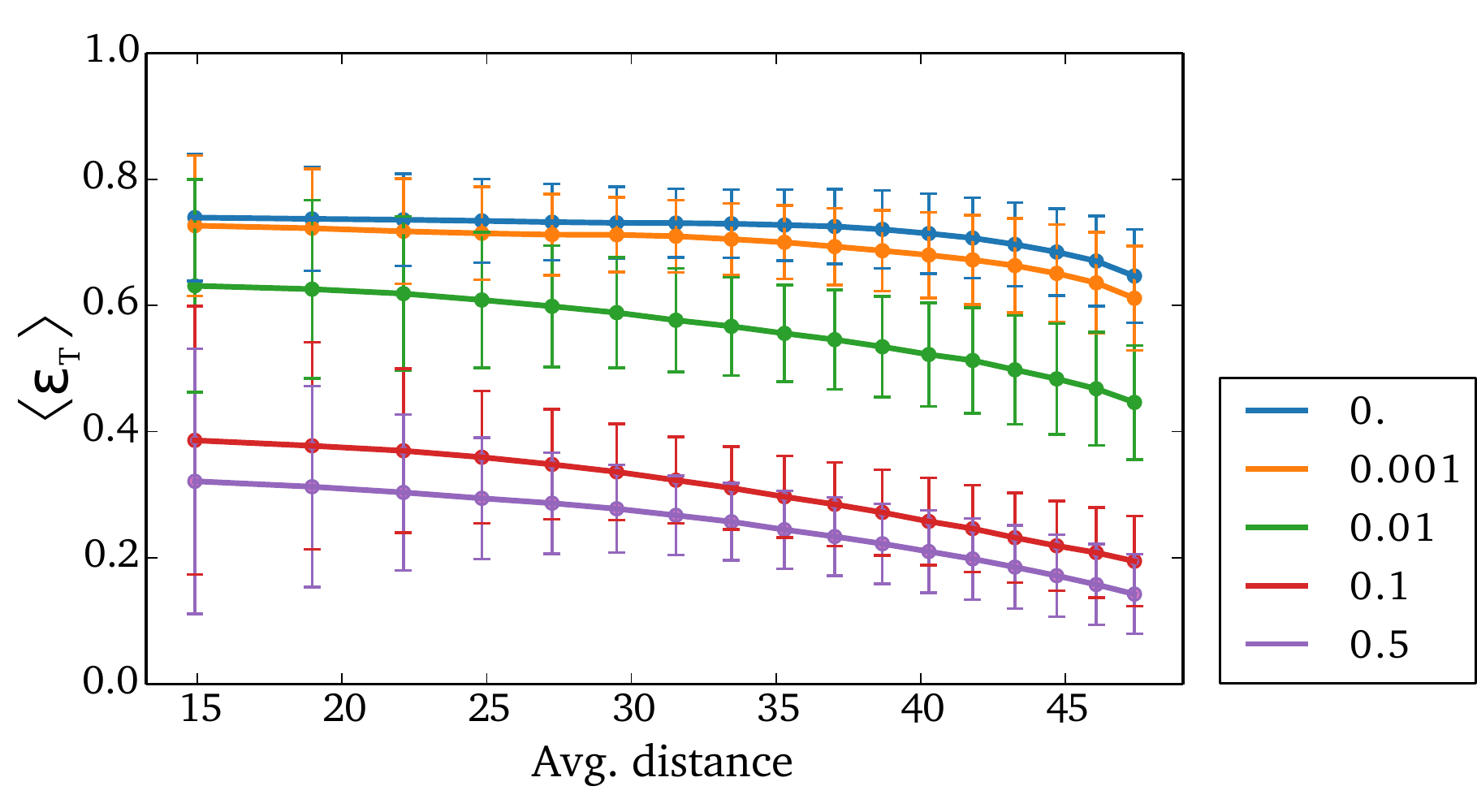}} \subfigure[$\tau$]{\includegraphics[width=0.85\linewidth]{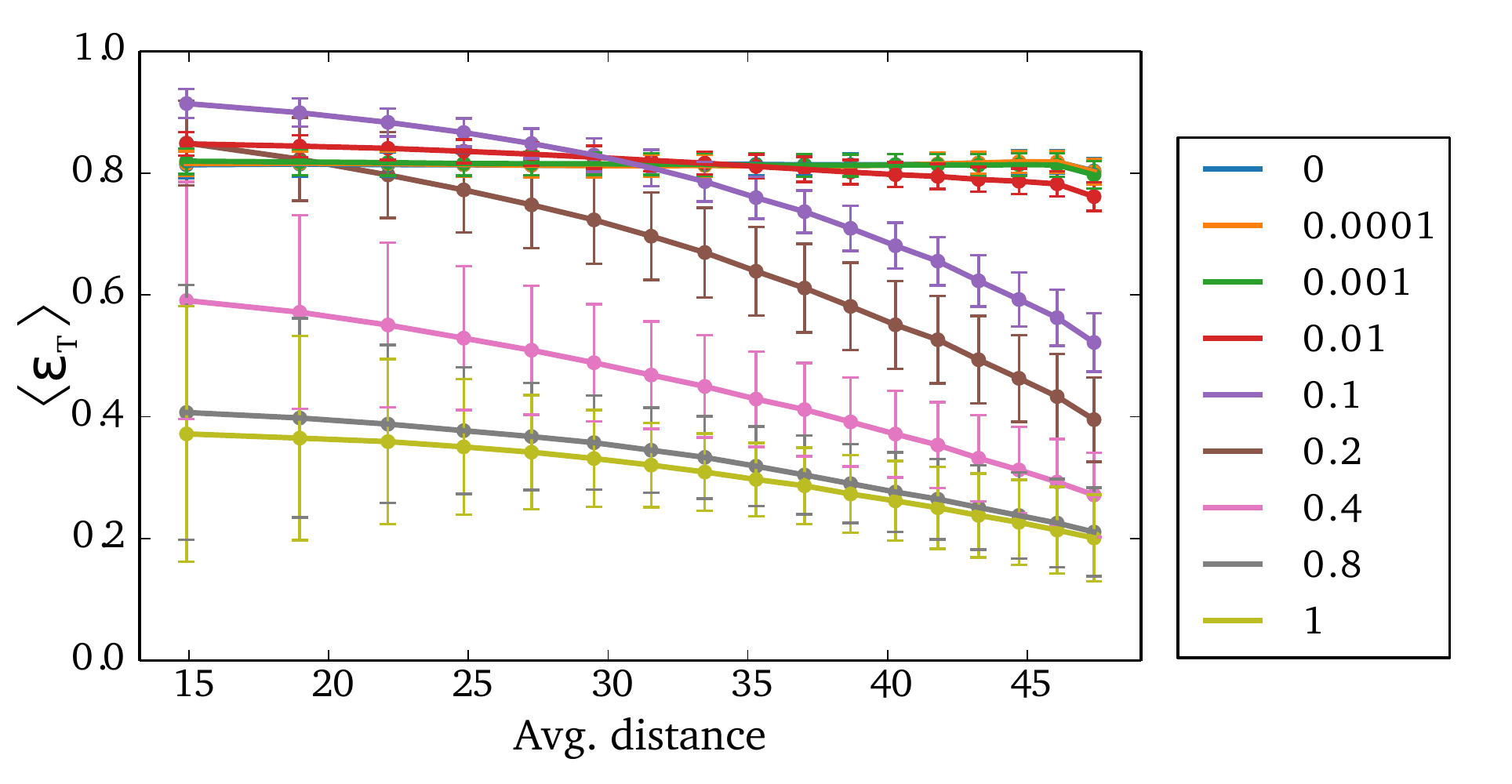}}
 \caption{The performance $\langle\varepsilon_T\rangle$ obtained from the dynamics simulation for each network region. Such regions were defined according to the distance between each node and the central position of LA. Two parameters, $\gamma$ (a) and $\tau$ (b), were varied.}
 \label{fig:fig_lattice_reg}
\end{figure}

The obtained results for the LA simulation can be seen in Figure~\ref{fig:fig_lattice_reg}, in which the performances of different $D_\eta$ values were not shown because, in this case, there are no significant variations among $\langle\varepsilon_T\rangle$ performance. The measured $\langle\varepsilon_T\rangle$ in different network regions was not found to present significant variations when the jump probability, $\gamma$, is changed (shown in Figure~\ref{fig:fig_lattice_reg}(a)). A similar effect was observed when a range of $D_\eta$ values was tested. Furthermore, as in the first analysis, varying the $\tau$ parameter provides a better performance for the regions near the geographical center of the network, but when $\tau \leq 0.01$ there were no significant variations of $\langle\varepsilon_T\rangle$ (shown in Figure~\ref{fig:fig_lattice_reg}(b)).

A more extensive analysis was done by considering the accessibility measurement, $\alpha^{(3)}$ to characterize the networks regions. According to this analysis, the variations of $D_\eta$ vector were distinctly reflected in the dynamics for each network region. Since the results obtained for the parameter $\tau$ are almost the same as those obtained for $D_\eta$, the following discussion regarding this parameter also holds for $D_\eta$.

\begin{figure*}[!htbp]
 \centering
 \subfigure[WS-2]{\includegraphics[width=0.32\linewidth]{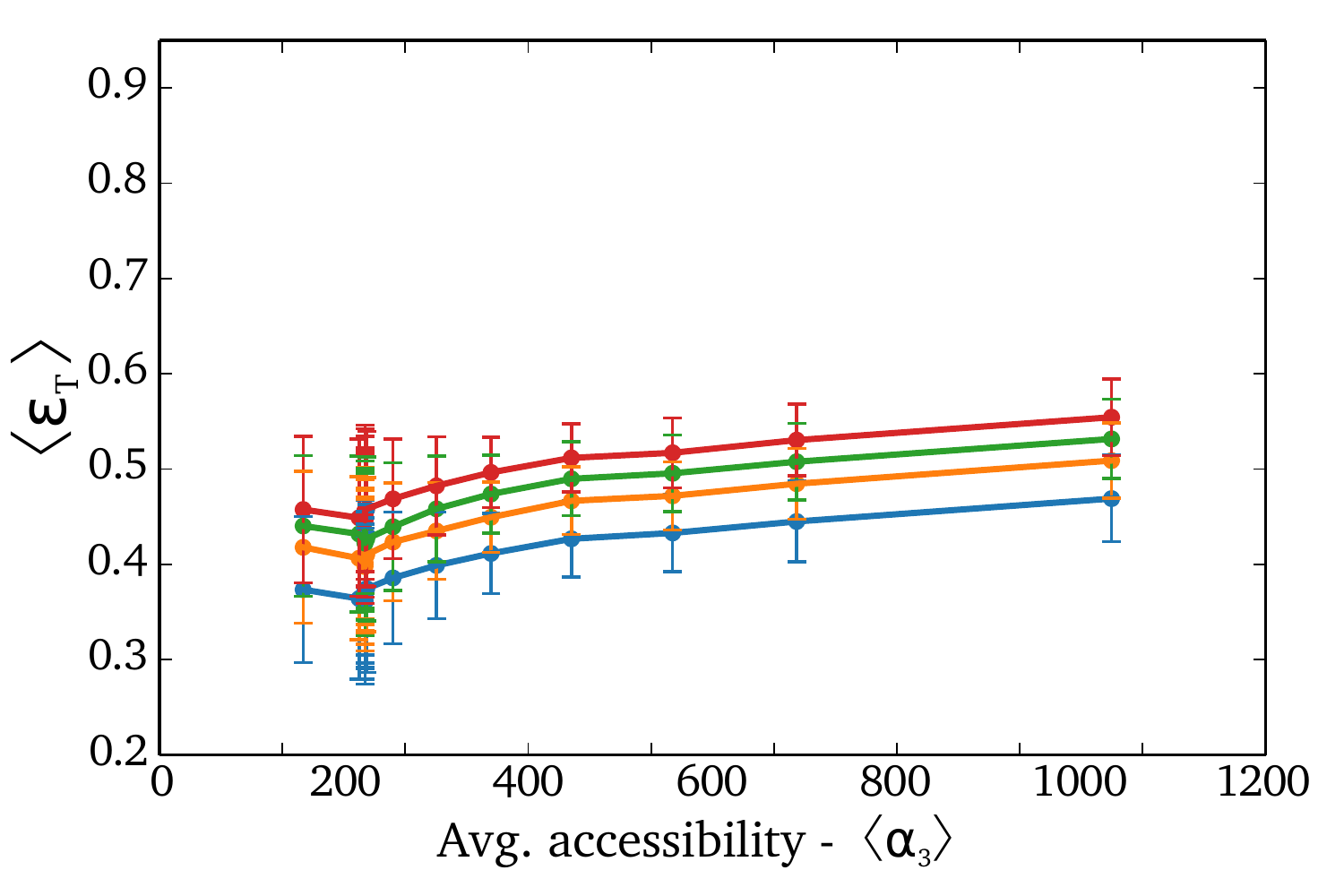}}
 \subfigure[WAX]{\includegraphics[width=0.32\linewidth]{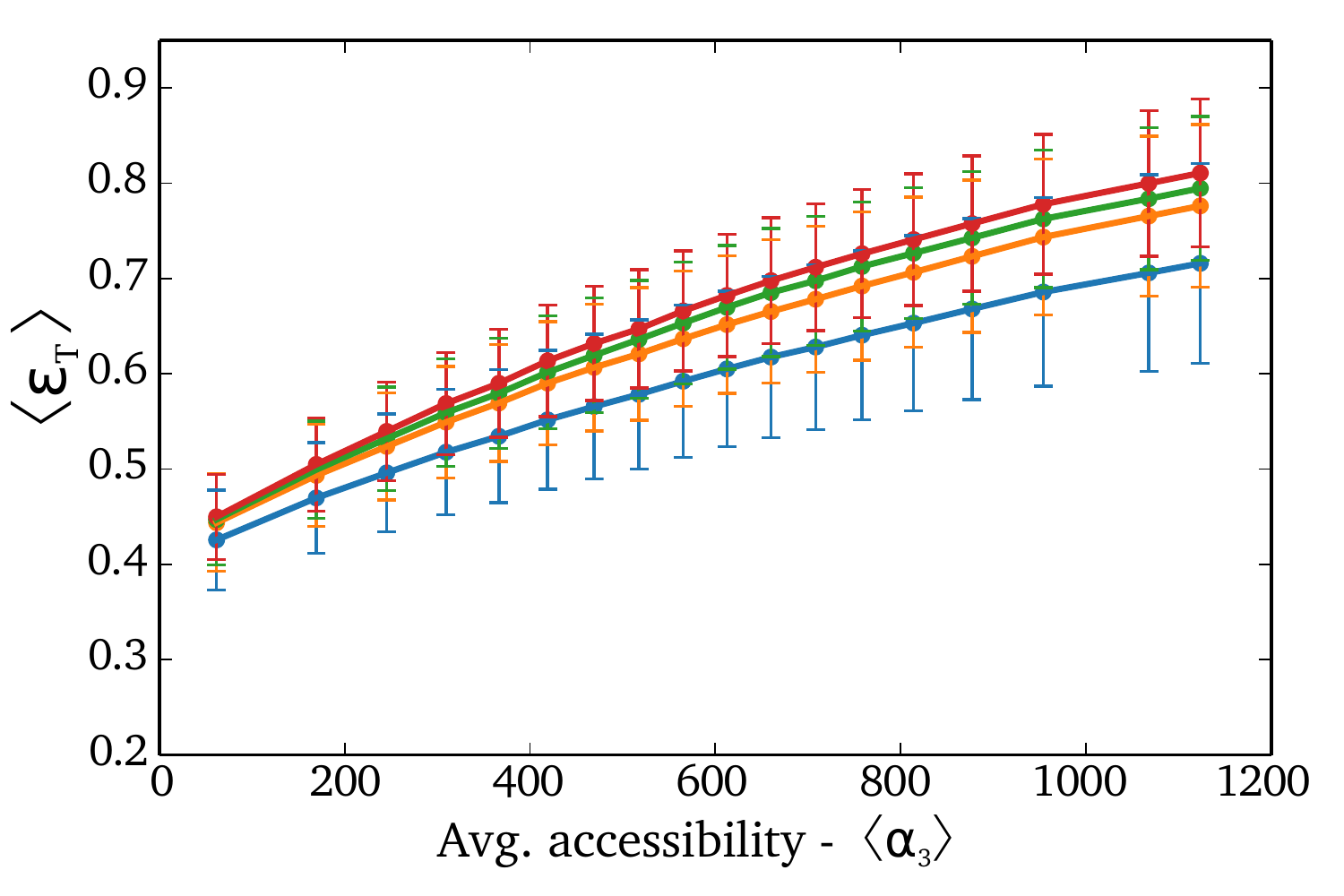}}
 \subfigure[CN]{\includegraphics[width=0.32\linewidth]{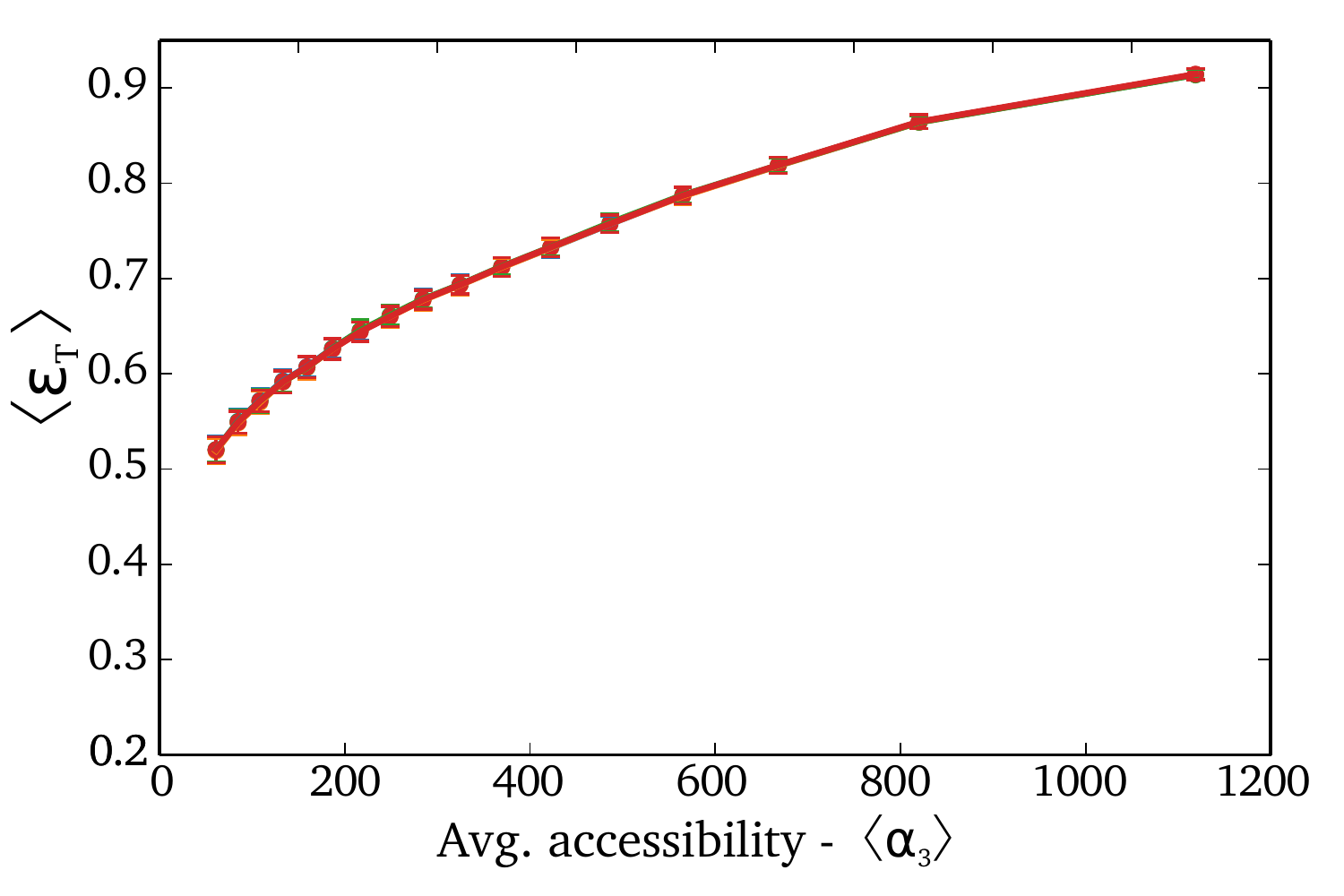}}\\
 \subfigure[WOS]{\includegraphics[width=0.32\linewidth]{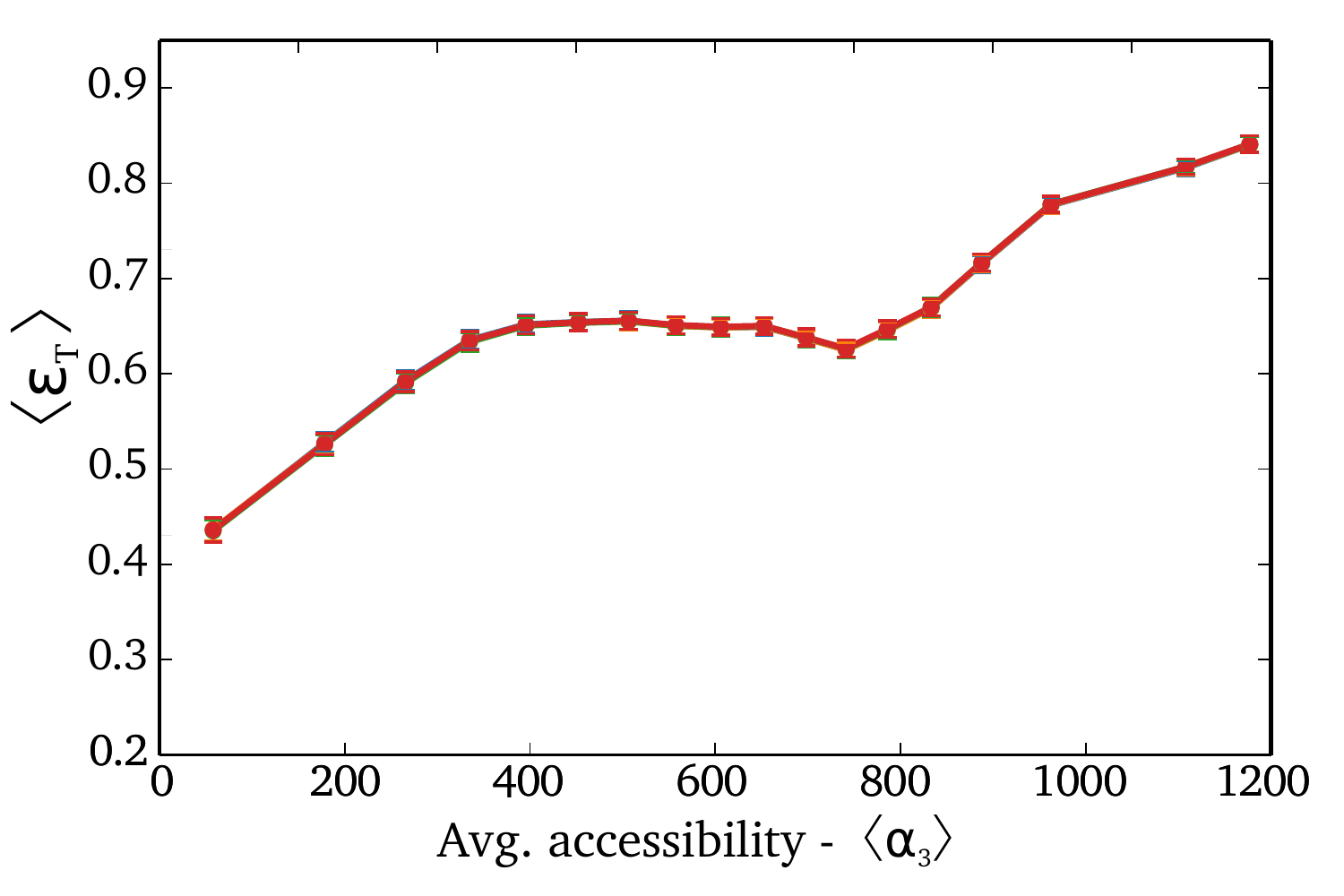}}
 \subfigure[WIKI]{\includegraphics[width=0.32\linewidth]{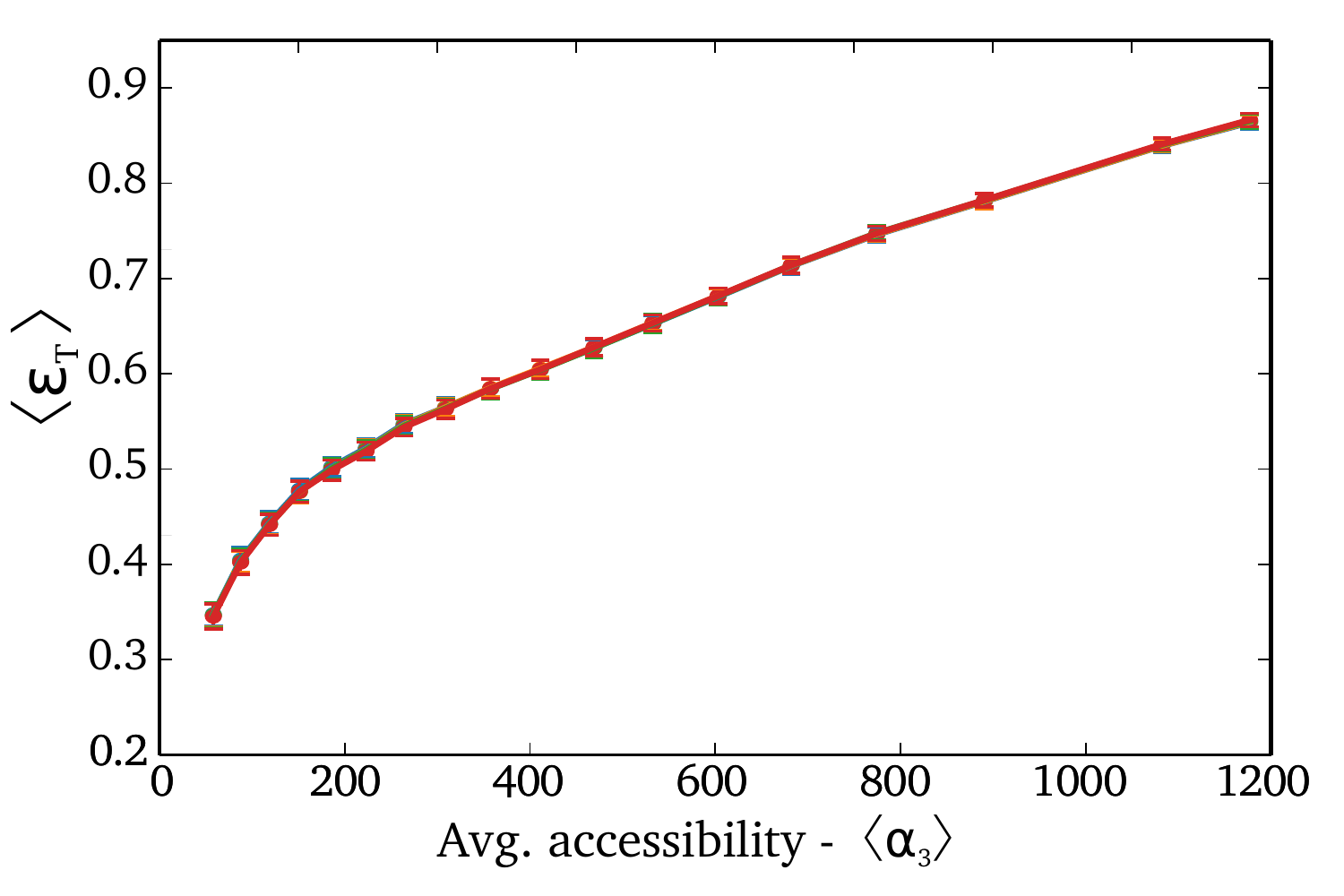}}
 \subfigure[BA]{\includegraphics[width=0.32\linewidth]{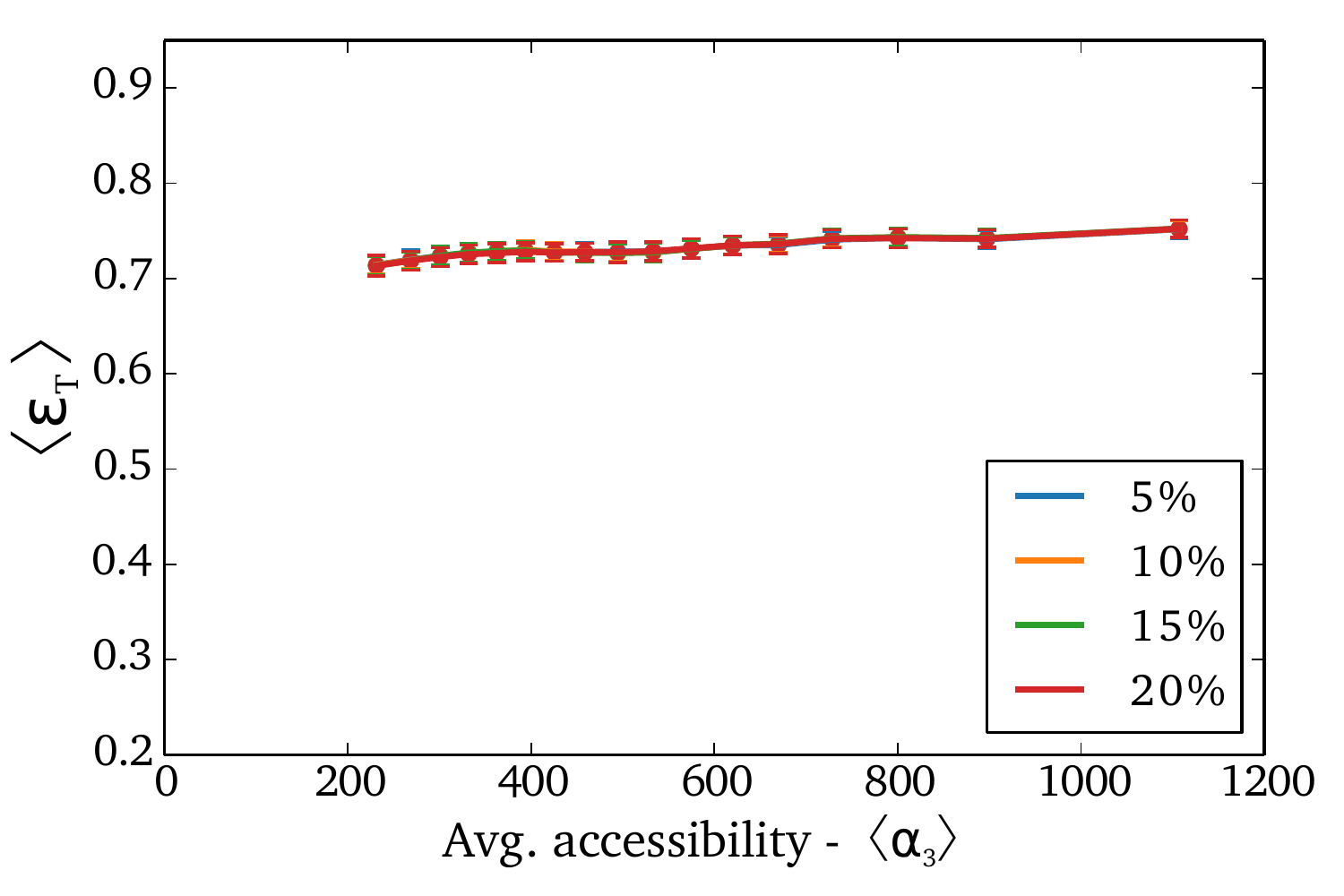}}
 \caption{The performance $\langle\varepsilon_T\rangle$ along the network regions for distinct values of $D_\eta$. The regions were defined in terms of the accessibility measurement with $h=3$. For this analysis, the considered networks are: (a)WS-2; (b)WAX; (c)CN, (d)WOS, (e)WIKI, (f)BA.}
 \label{fig:fig_eta_reg}
\end{figure*}

The performance obtained for the considered networks by varying $D_\eta$ are shown in Figure~\ref{fig:fig_eta_reg}. In most cases, variations of $D_\eta$ did not imply in performance variation for any of the network regions. However, when comparing among regions in the same network, the most central regions tended to have higher values of $\langle\varepsilon_T\rangle$. Such an effect was observed in all the considered networks representing knowledge structure (CN, WOS, WIKI and BA), as shown in Figures~\ref{fig:fig_eta_reg}(c-f). From these results, we conclude that the number of visible agents did not significantly impact on the network performance. The factor that most influenced the performance was the centrality of the region being explored, because normally the network center is easier to be explored. An exception to this trend is the BA network (f), which did not present significant performance variations when considering distinct network regions.

For the WAX model, we found that no significant differences of $\langle\varepsilon_T\rangle$ when $D_\eta$ is changed for regions at the borders of the network. However, more expressive differences of $\langle\varepsilon_T\rangle$ were observed in the central region. The standard deviations were also higher, as shown in Figure~\ref{fig:fig_eta_reg}(b).  The simulation for the WS networks showed that the percentage of influent agents, $D_\eta$, is directly related to $\langle\varepsilon_T\rangle$. These results are shown in Figure~\ref{fig:fig_eta_reg}(c).

\begin{figure*}[!htbp]
 \centering
 \subfigure[WS-2]{\includegraphics[width=0.32\linewidth]{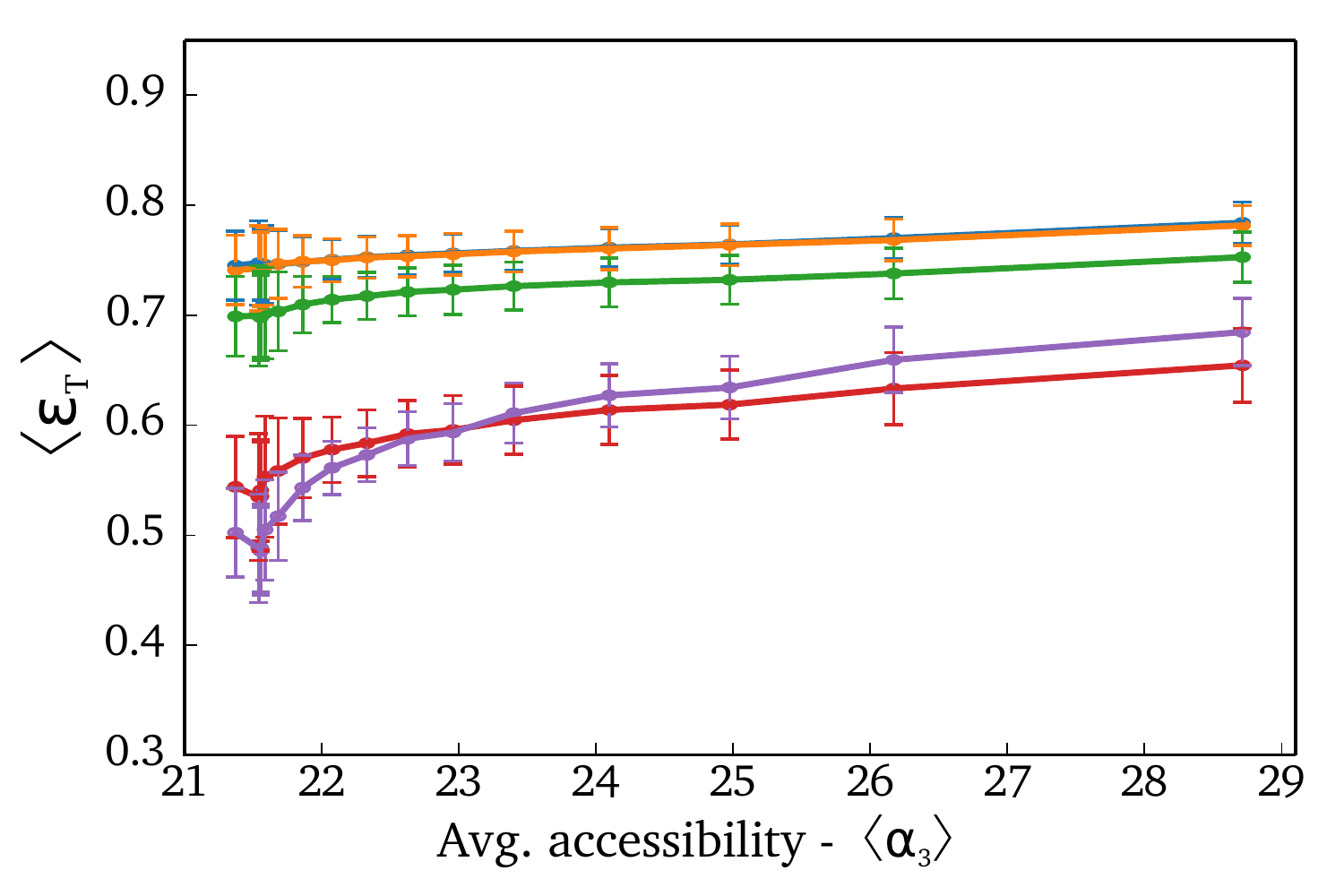}}
 \subfigure[WAX]{\includegraphics[width=0.32\linewidth]{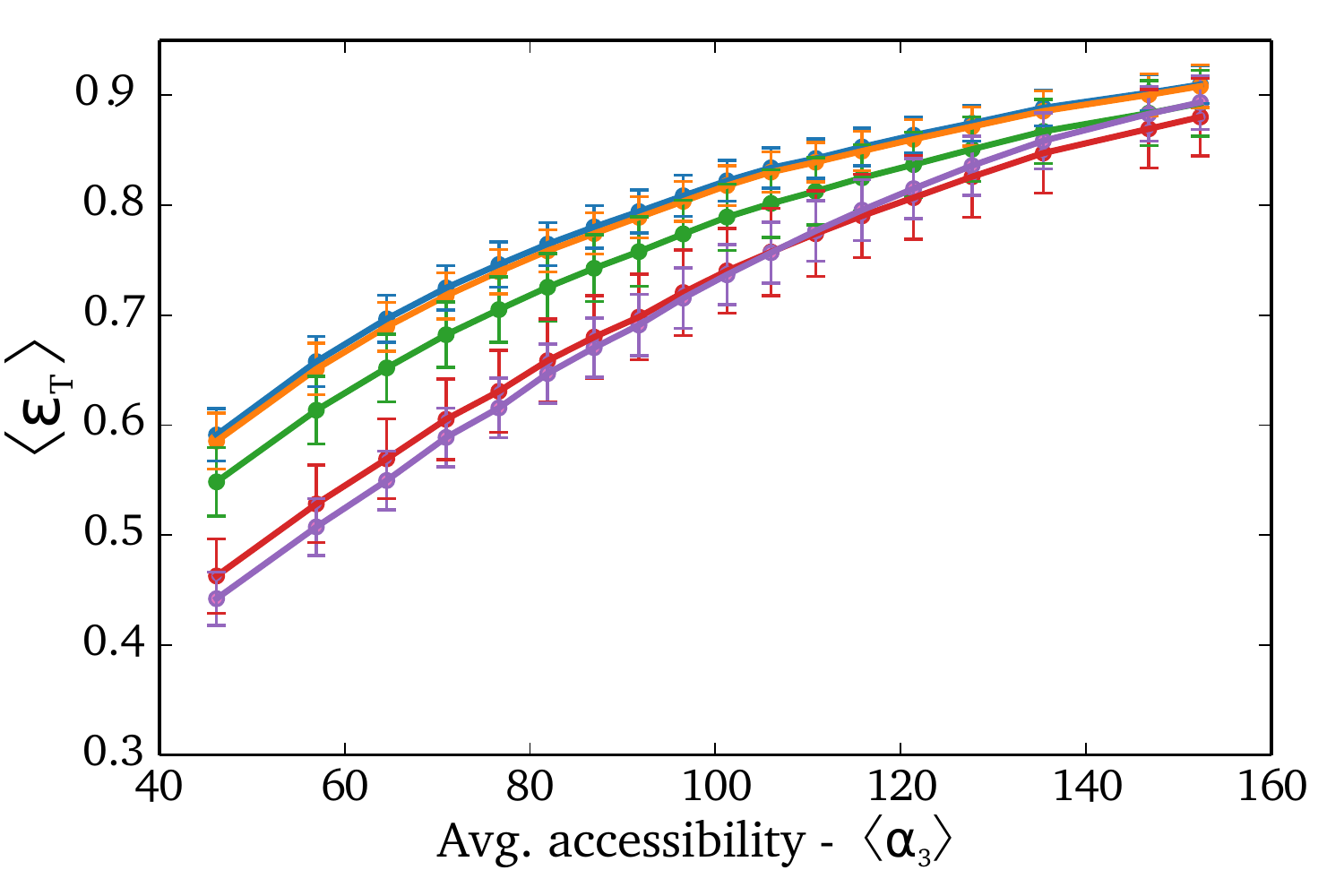}}
 \subfigure[CN]{\includegraphics[width=0.32\linewidth]{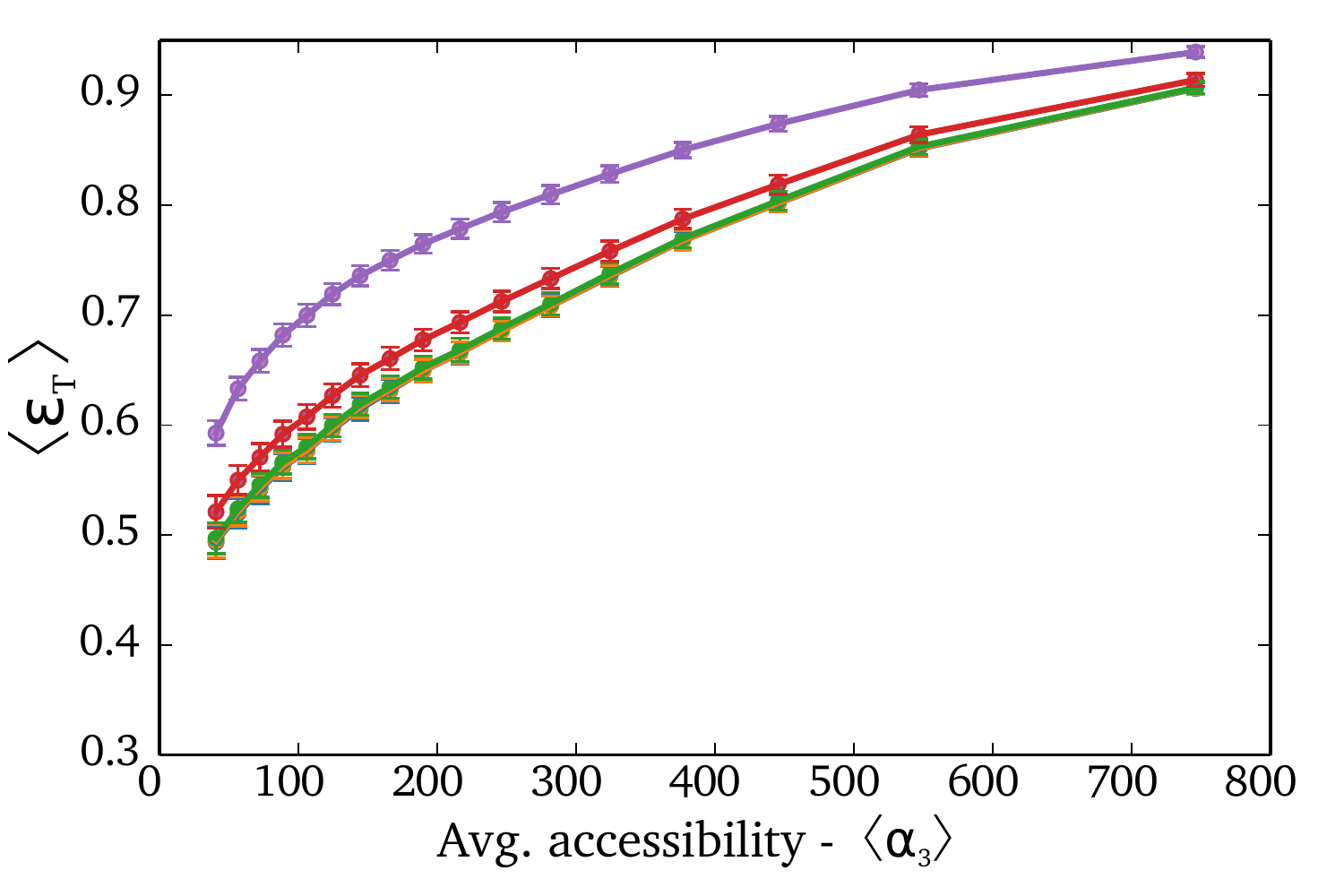}}\\
 \subfigure[WOS]{\includegraphics[width=0.32\linewidth]{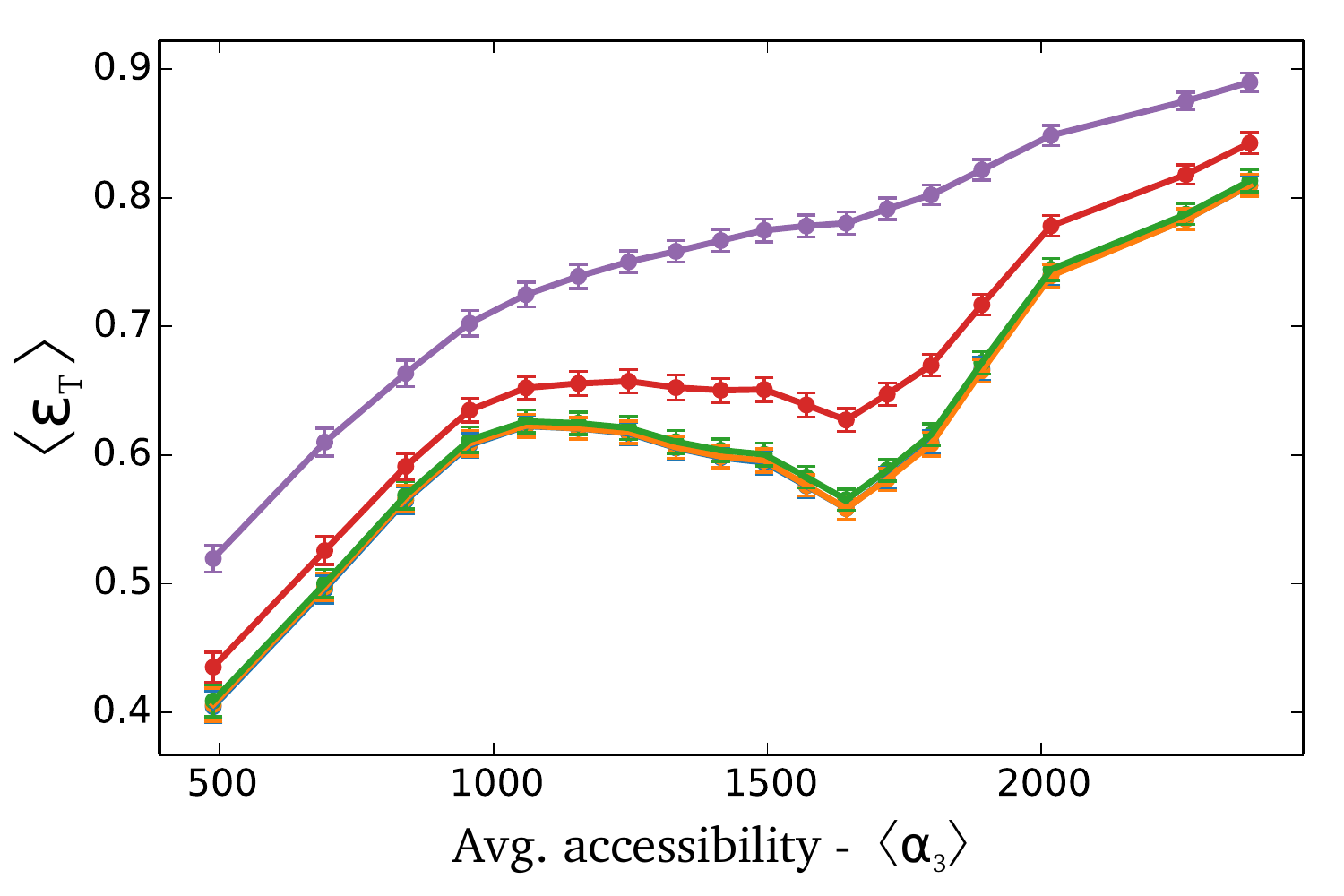}}
 \subfigure[WIKI]{\includegraphics[width=0.32\linewidth]{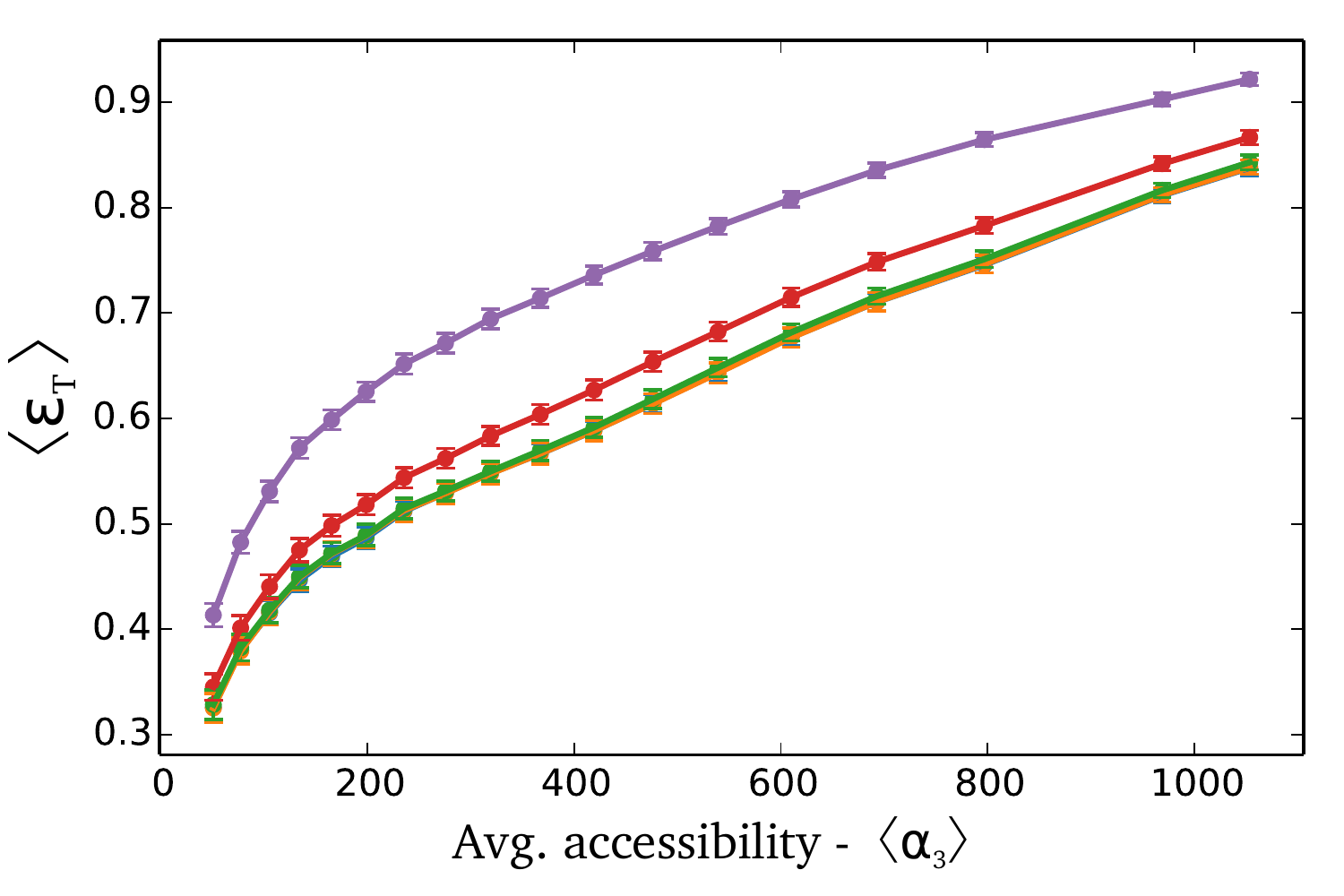}}
 \subfigure[BA]{\includegraphics[width=0.32\linewidth]{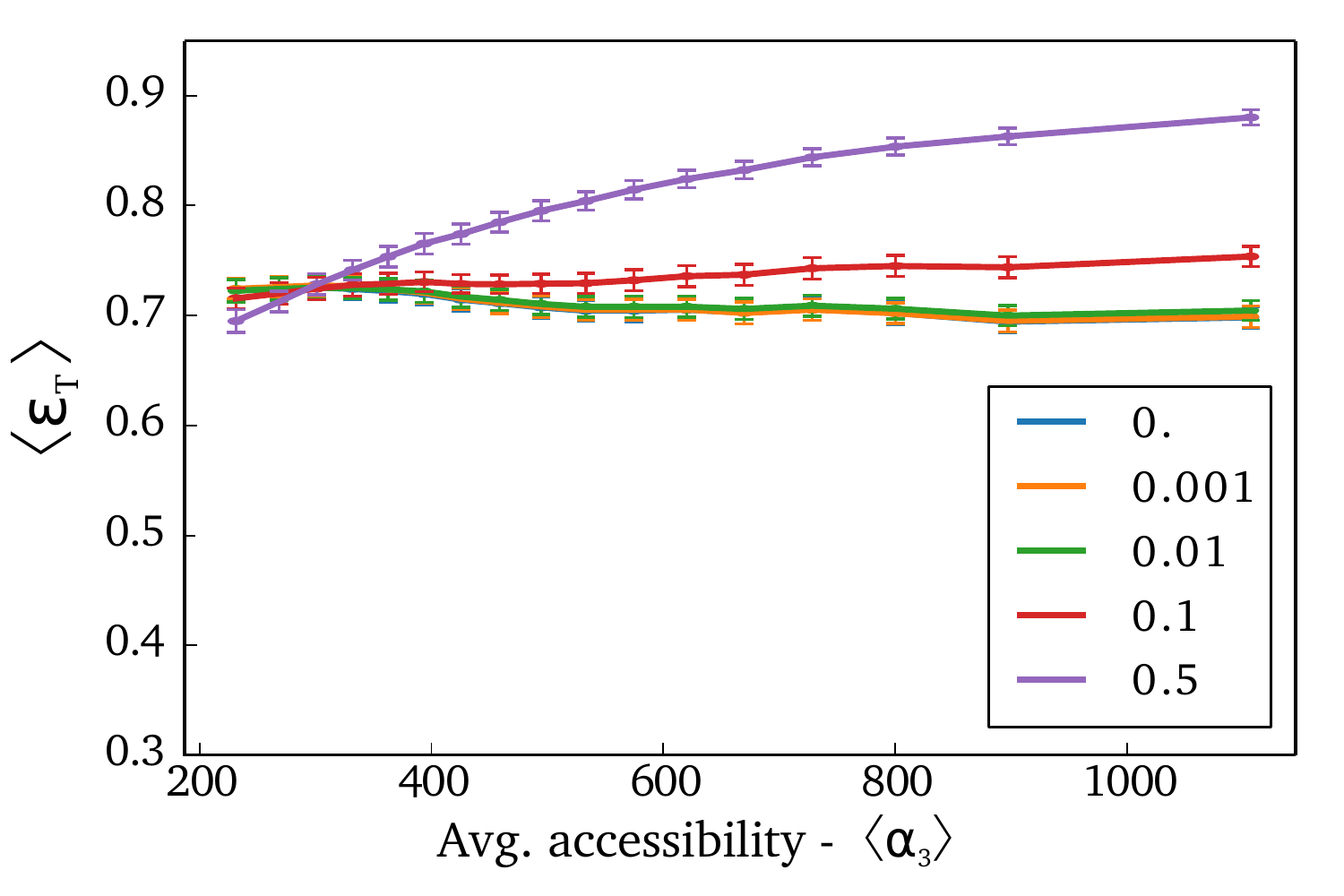}}
 \caption{The performance $\langle\varepsilon_T\rangle$ for the WS-2 (a), WAX (b), CN (c), WOS (d), WIKI (e) and BA (f) networks, measured from the dynamics simulation for each network region, where the regions were computed according to the $\alpha^{(3)}$ measurement. In this test, we vary the values of the parameter $\gamma$.}
 \label{fig:fig_jump_reg}
\end{figure*}

The results obtained for the analyses involving jump probability $\gamma$ variation are shown in Figure~\ref{fig:fig_jump_reg}. For all curves, the dynamics performance $\langle\varepsilon_T\rangle$ was found to be higher at the central regions of the networks (i.e. those with nodes having high accessibility). Except for the WOS(d) and BA(f) networks, the curves monotonically increase with the average accessibility of the regions. Similarly to the results obtained by varying $\eta$, the performance for the BA network does not vary substantially along the network regions nor by changing $\gamma$. However, it is interesting to note that, for regions presenting very low accessibility in the BA network, the performance has an opposite trend to the global behavior, as it decreases with $\gamma$.

In most cases, the influence of varying $\gamma$ to the performance tends to decrease with the average accessibility of the considered region. However, in the case of the WOS network, for a certain range of accessibility values around $1700$, the performance have a significantly drop of performance. This is interesting because other models did not present such behavior. Another interesting observation is that, the standard deviations obtained for the knowledge networks are substantially lower compared to those obtained for the spatial networks. This indicates that the peculiar behavior observed for the WOS networks is not caused by statistical fluctuations but may be the consequence of a more complex topological trait.

In general, for the considered networks representing knowledge structure (WOS, WIKI, CN and BA), in contrast with the spatial networks (LA, WS-2 and WAX), the dynamics do not change much by varying its parameters. In the average, the performance increases only by a small factor, for jump probability and changes even less for the other parameters. However, for the WOS network, depending on the network region, the improvement of performance by varying this parameter can be substantially higher than the other networks.

\section{Conclusion}

The problem of how humans represent and acquire knowledge has received growing attention along the last years as a consequence of its potential for understanding and improving the speed of research and learning (e.g.~\cite{foster2015tradition,cokol2005emergent}).  These problems are considerably complex because they involve several aspects that can influence the efficiency of how knowledge is achieved, exchanged and disseminated.  Such aspects include, for instance, the complexity of the knowledge structure itself, the visibility of researchers, the available memory, the strategies adopted for deciding the next possible subjects of research, amongst many others.  One particularly promising approach to understanding knowledge acquisition consists in representing the knowledge as a complex networks, and researchers as agents that move along such networks.  A number of approaches adopting such a framework have been proposed in the literature~\cite{doi:10.1093/comnet/cnu003,Pareschi20130396,batista2010knowledge} to study the learning process, i.e. the discovery of new concepts by researchers.  In this study, we propose a systematic approach focusing on two important elements influencing knowledge acquisition, namely the memory and visibility of agents, in order to better understand the collective discovery process.  The memory aspect was investigated  in terms of true self-avoiding dynamics occurring in a knowledge space. Additionally, interactions among researchers were modeled by means of a flight dynamics biased towards the most visible researchers.

We execute our dynamics in many networks, including two real networks (WIKI and WOS) and a set of network models with distinctive topological properties. We observed that the performance of knowledge acquisition can be distinctly optimized for different characteristics of the networks. For example, by increasing the jump probability ($\gamma$), the speed of learning also rises significantly in the case of BA networks. This could be a consequence of the fact that, in such networks, the agents need to pass through hubs to access other nodes up to the point that it starts to avoid the hubs completely. This results in such an agent becoming trapped by a set of nodes because the majority of the shortest paths go across hubs. In a BA network, a jumping agent can move without crossing the hubs, thus having more possibilities to access new nodes. On the other hand, such a configuration was found to be less effective in WAX networks, since when an agent jumps towards other nodes, it will tend to navigate through already learned concepts.

We also investigated the performance by considering distinct network regions, such as the borders and the center of networks. We found that, in most cases, collective discovery occurs faster at the core of the network and becomes slower at the borders. Regarding the three parameters controlling the dynamics, in contrast with the results obtained for the global analysis, some dynamics configurations can indeed change the learning performance depending on the properties of regions being explored. However, for the networks representing knowledge structure, the average gain in performance is low irrespectively to the dynamics parameters.  An exception to this trend is the WOS network, which presents a set of regions where the performance can vary substantially under influence of the jump probability. In this case, it is possible to enhance knowledge acquisition by a substantial factor. These results indicate that, for a typical knowledge network, the heuristics governing the way researchers seek for new knowledge does not substantially affect the global performance of the collective discovery, but still can have influence depending on the properties of the region of knowledge under investigation.

The current study adopted a random walk approach incorporating jumps in a multi-agent dynamics. In future works, we intend to probe the effects of considering other network topologies, including directed structures, on the proposed collective learning model. Additionally, some dynamic characteristics can also be investigated, such as other visibility mechanisms and agents with different features (e.g. memory, speed, etc.). Furthermore, our study can be extended to incorporate two layers: one representing the knowledge organization, and the other a network of interactions among researchers.

\section*{Acknowledgements}
The authors acknowledge financial support from Capes-Brazil,  S\~ao Paulo Research Foundation (FAPESP) (grant no. 2016/19069-9,
2015/08003-4, 2014/20830-0 and 2011/50761-2), CNPq-Brazil (grant no. 307333/2013-2) and NAP-PRP-USP.

\bibliography{manuscript}

\end{document}